\documentclass [11pt]{article}
\usepackage{setspace}
\usepackage{amsfonts}
\usepackage{amsthm}
\usepackage{color}
\usepackage{graphicx}
\usepackage{rotating}

\usepackage{verbatim}

\newcommand\ba {\mathbf a}
\newcommand\bb {\mathbf b}

\newcommand\bd {\mathbf d}

\newcommand\bm {\mathbf m}

\newcommand\bt {\mathbf t}
\newcommand\bu {\mathbf u}
\newcommand\bv {\mathbf v}

\newcommand\bx {\mathbf x}
\newcommand\by {\mathbf y}
\newcommand\bz {\mathbf z}

\newcommand\bA {\mathbf A}

\newcommand\bI {\mathbf I}

\newcommand\bS {\mathbf S}

\newcommand\bU {\mathbf U}
\newcommand\bV {\mathbf V}
\newcommand\bW {\mathbf W}
\newcommand\bX {\mathbf X}
\newcommand\bY {\mathbf Y}

\newcommand\bZ {\mathbf Z}

\newcommand\indica {\mathbb{I}}

\newcommand\wbd {\widehat{\bd}}

\newcommand\itA {{\mathcal{A}}}
\newcommand\itB {{\mathcal{B}}}

\newcommand\itE {{\mathcal{E}}}
\newcommand\itF {{\mathcal{F}}}
\newcommand\itG {{\mathcal{G}}}

\newcommand\itI {{\mathcal{I}}}

\newcommand\itM {{\mathcal{M}}}

\newcommand\itS {{\mathcal{S}}}
\newcommand\itT {{\mathcal{T}}}
\newcommand\itU {{\mathcal{U}}}

\newcommand\itW {{\mathcal{W}}}

\newcommand\bbe {\mbox{\boldmath $\beta$}}

\newcommand\bbech {\mbox{\scriptsize${\bbe}$}}

\newcommand\bmu {\mbox{\boldmath $\mu$}}
\newcommand\bmuch {\mbox{\scriptsize\boldmath $\mu$}}

\newcommand\bSi {\mbox{\boldmath $\Sigma$}}
\newcommand\bSich {\mbox{\scriptsize\boldmath $\Sigma$}}

\newcommand\wbmu {\widehat{\bmu}}

\newcommand\wbSi {\widehat{\bSi}}

\newcommand{\ml}{\mbox{\scriptsize \sc cl}}
\newcommand{\ese}{\mbox{\scriptsize \sc s}}
\newcommand{\ds}{\mbox{\scriptsize \sc ds}}
\newcommand{\bat}{\mbox{\scriptsize \sc bat}}
\newcommand{\sch}{\mbox{\scriptsize \sc sch}}

\def\real{\mathbb{R}}

\def\qu{\mathbb{Q}}
\def\complejo{\mathbb{C}}

\newcommand{\esp}{\mathbb{E}}
\newcommand{\prob}{\mathbb{P}}

\newcommand{\var}{\mbox{\sc Var}}

\newcommand{\convpp}{ \buildrel{a.s.}\over\longrightarrow}

\newcommand{\convprob  }{ \buildrel{p}\over\longrightarrow}

\newcommand{\convdist}{ \buildrel{D}\over\longrightarrow}

\newcommand{\trasp}{^{\mbox{\footnotesize \sc t}}}

\newcommand\bcero {{\bf{0}}}

\newcommand{\identidad}{\mbox{\bf I}}

\newcommand{\IF}{{\mbox{IF}}}

\newcommand\noi{\noindent}

\def\square{\ifmmode\sqr\else{$\sqr$}\fi}
\def\sqr{\vcenter{
         \hrule height.1mm
         \hbox{\vrule width.1mm height2.2mm\kern2.18mm
\vrule width.1mm}
         \hrule height.1mm}}

\usepackage{color}

\begin{document}

\title{Conditional tests for elliptical symmetry using robust estimators}
\author{Ana M. Bianco\\
 \footnotesize Facultad de Ciencias Exactas y Naturales, Universidad de Buenos Aires and CONICET\\
Graciela Boente
\\ \footnotesize Facultad de Ciencias Exactas y Naturales, Universidad de Buenos Aires and CONICET\\
Isabel M. Rodrigues \\
\footnotesize Departamento de Matem\'atica and CEMAT, Instituto Superior T\'ecnico, Universidade de Lisboa}

 \date{}
\maketitle

\begin{abstract}
 This paper presents a procedure for testing the hypothesis that the underlying distribution
of the data is elliptical when using robust location and scatter estimators instead of the sample
mean and covariance matrix. {Under mild assumptions that  include elliptical distributions without first moments, we derive the test statistic asymptotic behaviour under the null hypothesis and under special alternatives.} Numerical experiments allow to compare the behaviour of the tests based on the sample mean and covariance matrix with that based on robust estimators, under various elliptical distributions  and different alternatives. This comparison was done looking not only at the observed level and power but we rather use the size--corrected relative exact power which provides a tool to assess the test statistic skill to detect alternatives. We also provide a numerical comparison with other competing tests.

\vskip0.2in
\noi\textbf{Key Words:} {Bootstrap, Conditional test, Elliptical symmetry, Robust estimation }

\noi\textbf{AMS Subject Classification:} {MSC 62H15; 62F35; 62F40}

\end{abstract}

\normalsize
\newpage

\section{Introduction}{\label{intro}}

The family of elliptically symmetric (or elliptically contoured) distributions  generalizes the family of multivariate normal distributions. One advantage of the elliptical distributions is that they define a much broader class of multivariate distributions than the multivariate normal distributions  so that they can serve as the basis for the development of more robust analyses. In fact, in many situations, normal--theory analyses can be modified slightly retaining their validity across all elliptical distributions.  
The fact
that many statistical procedures (including principal component analysis)
yield superior performance when data support elliptical symmetry, motivates the consideration of testing
for elliptical symmetry, instead of testing for other forms of multivariate
symmetry.

 Zhu and Neuhaus (2003) introduced   conditional test procedures for testing elliptical symmetry of a multivariate distribution. The conditional tests are exactly valid if the center and the shape matrix are known and are asymptotically valid if they are estimated,  when fourth moments exist. It is worth noting that the test proposed by Zhu and Neuhaus (2003) are based on the sample mean and the sample covariance matrix, when the center and/or the shape parameters are unknown. This entails that the test statistics are asymptotically  valid only for elliptical distributions such that $\esp (\|\bX\|^4)<\infty$. In a robust framework, one frequently assumes that the sample belongs to a neighbourhood of a given central elliptical distribution $P_0$. The distributions $P$ to be considered in the neighbourhood include heavy tailed distributions. Furthermore, in order to ensure Fisher--consistency of the proposed estimators, it is generally assumed that the resulting distribution $P$ is also elliptical. So,   it is of interest to check if the assumption of elliptical symmetry is valid without making moment  assumptions. 
For this reason, in this paper, we propose a testing procedure that can be helpful to decide if a given sample has a common elliptical distribution without requiring moment conditions when consistent estimators of the unknown parameters are available.

The paper is organized as follows. In Section \ref{test}, we introduce our proposal, while  asymptotic distribution results under the null and under contiguous alternatives are provided in Section \ref{main}. A bootstrap method to compute effectively an approximation of the proposed test is presented in Section \ref{boot}.  The results of a Monte Carlo study in dimensions $p=2$ and $5$ are summarized in Section \ref{monte}, while a procedure to compute the test statistic is described in   Appendix A. Besides, in Section \ref{numerical},    we   analyse the behaviour of the proposed test statistic and the classical one under different distributions and sample sizes, so as to check their ability to reject the null hypothesis against a set of alternatives. Proofs are relegated to Appendix B and C.

\section{Test statistic}{\label{test}}
For the sake of completeness, we briefly recall the notion of elliptical symmetry. One can define symmetry in terms of structural properties of the distribution function, of the density function or of the characteristic function.
The distribution of a $p$-dimensional random vector $\bY$ is called
 spherically symmetric when, for every $p \times p$ matrix  $\bA \in  {\cal O}(p)$
(the orthogonal group), the distribution of $\bA \bY$ is the same as that of $\bX$.
A random vector $\bX \in \real^p$  has an elliptically symmetric or elliptically contoured
distribution,  with parameters $\bmu \in \real^p$ and  a non-singular matrix $\bSi$, if  $ \bZ=\bSi^{-1/2} (\bX-\bmu)$ is  a spherically
symmetric random vector. If this elliptical distribution has finite second moments, then $\bmu$ is the mean vector and $\bSi$ is up to a scalar the covariance matrix. More generally, under no moment conditions, the parameters $\bmu$ and $\bSi$ are called the location and the scatter matrix parameters, respectively.
The associated characteristic function of an elliptical vector has the form
$ 
\phi(\bt)=e^{i \bt\trasp \bmu} \psi (\bt\trasp\bSi \bt)$,
for $\bt \in \real^p$, for some
scalar function $\psi:\real \to \real$. Then,  if second moment exists $\var(\bX)=\,-\,2\psi'(0)\bSi$, where for identifiability of $\bSi$ it is usually required that $\psi'(0)={\,-\,}1/2$. We will write that $\bX \sim \itE_p(\bmu, \bSi, \psi)$. 
If $\psi(x) = e^{-{x}/{2}}$ in the expression of $\phi(\bt)$, we get the characteristic function of a normal distribution, so  elliptical distributions are generalizations of normal
distributions. For an overview on these distributions we refer to Fang and Anderson (1990) and Fang \textsl{et al.} (1990).

 Among the tests for spherical and elliptical distributions that have been introduced, we can mention Beran (1979), Tyler (1982), Baringhaus (1991), Fang \textsl{et al.} (1993), Koltchinskii and Li (1998), Koltchinskii and Sakhanenko (2000), Schott (2002), Zhu and Neuhaus (2003),  Huffer and Park (2007) and more recently, Batsidis and Zografos (2013) and Batsidis \textsl{et al.} (2014). 

The goal of this section is to suggest a modification of the conditional test for ellipsoidal symmetric multivariate distributions proposed in Zhu and Neuhaus (2003), which allows its application to data coming from heavy tailed elliptical distributions. This is particularly appealing in a robust framework, since many resistant statistical procedures assume that the underlying distribution is elliptical to get Fisher--consistent estimators.

Let $\bX$ be a $p$-dimensional random vector with distribution $P$.  Given independent and identically distributed (i.i.d.) observations $\bX_1, \dots, \bX_n$ such that $\bX_i\sim \bX$, denote by $P_n$  the empirical measure based on the sample points. Moreover, let $P_n f=P_n(f)$ stand for $(1/n)\sum_{j=1}^{n}f(\bX_j)$ for any function $f:\real^p\to \real$.

From now on, denote $\itE_p$ the class of all elliptical contoured distributions. The hypothesis to be tested is $H_0: P\in \itE_p$, that is, 
$
H_0: \bX \sim  \itE_p(\bmu, \bSi, \psi)$,
with parameters $\bmu$ and  $\bSi$. For each fixed $\bb$ and $\bA$ define the functional
\begin{equation}
 T_{\bb,\bA}(P)=\int_{\itS_p} \int   \left[\esp_P\left( \sin \left \lbrace t\ba\trasp \bA^{-1/2} \left( \bX-\bb \right) \right\rbrace \right)  \right]^2\; w(t)dtdv(\ba)
 \label{funcional}
 \end{equation}
 where   $w:\real\to \real$ is a weight function , $\itS_p= \lbrace \ba \in \real^p : \|\ba\|=1\rbrace$, $v$   the uniform distribution on $\itS_p$ and  $\esp_P$ indicates that the expectation is taken with respect to the probability measure $P$.
 Note that if $\bX \sim  P=\itE_p(\bmu, \bSi, \psi)$, then $\bZ=\bSi^{-1/2}(\bX-\bmu)$ is spherically distributed. Hence, the imaginary part of its characteristic function vanishes, that is, $\esp_{ P} \left[\sin(t\ba\trasp \bZ)\right]=0$ for any $t\in \real$ and $\ba \in  {\itS_p}$ which implies that $T_{\bmuch,\bSich}(P)=0$. 
 
 When $\bmu$ and $\bSi$ are known, the empirical version of $T_{\bmuch,\bSich}(P)$ will also be close to zero. This suggests to reject the null hypothesis $H_0$  for large values of the test statistic $T_n(\bmu, \bSi)$, where 
\begin{equation}
T_n(\bb, \bA)=\int_{\itS_p} \int  \left \lbrace \sqrt{n}P_n \sin\left[ t\ba\trasp \bA^{-1/2} \left( \bX-\bb \right) \right]  \right \rbrace ^2\, w(t)dtdv(\ba)\,.
\label{clatest1}
\end{equation} 
The statistic  $T_n(\bmu, \bSi)$ was considered by Zhu and Neuhaus (2003) when $w:\real\to \real$ has a  compact  support $\itI $ and is a weighted version of that studied  by Ghosh and Ruymgaart (1992).  

Usually the location and scatter matrix parameters are unknown. To overcome this problem, one may replace in $T_n(\bmu, \bSi)$, $\bmu$ and $\bSi$ by consistent estimators. 
Zhu and Neuhaus (2003) suggested to use the   classical (\textsc{cl}) sample estimators, leading to the  test statistic  $T_{n,\ml}=T_n(\wbmu, \wbSi)$, 
where $\wbmu=\bar{\bX}=\sum_{i=1}^n \bX_i/n$  and $\wbSi=\bS=\sum_{i=1}^n (\bX_i-\bar{\bX})(\bX_i-\bar{\bX})\trasp/n$. As mentioned in  Anderson \textsl{et al.} (1986), for a sub-class of elliptical distributions,  the maximum likelihood estimator  of $\bmu$ is $\bar{\bX}$, while that of   $\bSi$ is a constant multiple (depending on the family) of the sample covariance matrix, that is, the estimators have the same form as in the normal case, which justifies the above choice.  {Anderson \textsl{et al.} (1986) studied the general situation of elliptical random matrices $\mathbb{X}=(\bX_1, \dots, \bX_n)$ which includes the   setting of independent columns we are considering. In the particular case of independent random vectors $\bX_i$, the class of elliptical distributions for which  the sample mean  and   covariance matrix (except for a constant) are still the maximum likelihood estimators includes the}   situation   when  the density of $\bX_1$ equals $\det(\bSi)^{-1/2}\,g((\bx-\bmu)\trasp\bSi^{-1}(\bx-\bmu))$ for some $g:\real \to \real^{+}$, where $y^{p/2} g(y)$ has a finite positive maximum. A condition ensuring the existence of finite positive maximum is the continuity of $g$  and that $\esp \, \|\bX\|^2<\infty $  (see Lemma 2 in Anderson \textsl{et al.}, 1986). On the other hand, if the underlying distribution is heavy tailed the values of these estimators 
may be distorted,  rendering meaningless the test results.  A solution to this problem is well known in robust statistics: $\bmu$ and $\bSi$ have to be estimated in a robust manner, to provide consistent estimators even if moments do not exist as in the case of a multivariate Cauchy distribution.

 The proposal in this paper, consists in plugging into the conditional  test statistic $T_n(\bmu, \bSi)$ robust consistent estimators $\bm_n$ and $\bV_n$ of the location $\bmu$ and scatter matrix $\bSi$, respectively, to test $H_0: P\in \itE_p$. This leads to the following conditional robust based statistic 
$$
T_{n,\bm,\bV}=T_n(\bm_n,\bV_n)=\int_{\itS_p} \int  \left \lbrace \sqrt{n}P_n \sin\left[ t\ba\trasp \bV_n^{-1/2} \left( \bX-\bm_n \right) \right]  \right \rbrace ^2\, w(t)dtdv(\ba)\,,$$ 
where $P_n$, $\ba$, $w$ and $v$  are defined as in  (\ref{clatest1}).
Denote $\bV(P)$ and $\bm(P)$ the functionals related to $\bV_n$ and $\bm_n$, respectively, when   $\bX\sim P$. Usually, under $H_0$, $\bm(P)=\bmu$ and $\bV(P)$ is,  up to a multiplicative constant,  equal to $\bSi$. Then, the   functional related to $T_n(\bm_n,\bV_n)$ is just  $
T_{ \bm(P),\bV(P) }(P)$ defined in (\ref{funcional}) which justifies the considered procedure. 
In Appendix \ref{app}, we describe a numerical procedure to compute this test statistic.
 
\vskip0.1in
\noi \textbf{Remark \ref{test}.1.} The results on characteristic functions given in  Ushakov (1999) give some insight with respect to the choice of the weight function $w$. Indeed, as defined in Ushakov (1999) a characteristic function $\Phi$ is said to be analytic if there exists a function $g:\complejo\to \complejo$ which is analytic in $\{|z|\le R\}$ for some $R>0$ and such that $\Phi(t)=g(t)$ for any $t\in [-R,R]$.   Theorem 1.7.7 in Ushakov (1999) states that  if a characteristic function  $\varphi$ coincides with an analytic characteristic function $\Phi$ in some real neighbourhood of the origin, then  they coincide for all real, that is, $\varphi=\Phi$.

 Given a random vector $\bX\in \real^p$, denote   $Z_{\ba}$ the random variable $Z_{\ba}=\ba\trasp\bSi^{-1/2}(\bX-\bmu)$ and   $\varphi_\ba$ its characteristic function. Assume that for any $\ba \in \itS_p$, $\varphi_\ba$ is an analytic characteristic function.  Theorem 1.7.7 in Ushakov (1999)  entails that if,  for any $\ba \in \itS_p$, and for some $\delta>0$ we have that $\varphi_\ba(t)= \mbox{Re}(\varphi_\ba)(t)$ for $t\in (-\delta,\delta)$ then, the random variable $Z_{\ba}$ has a symmetric distribution for all $\ba$.  So, in the situation where the underlying distribution is such that all the projections have analytic characteristic functions,   which includes the multivariate normal  as well as the uniform    distribution on the ball or in sphere, if the functional related to the test statistic is zero for some weight function $w$, with support around $0$, then it will be 0 for any weight function. Thus, for   probability measures such that for all $\ba\in \itS_p$, the distribution of $Z_{\ba}$ has an analytic characteristic function the choice of $w$ is not crucial as far as its support contains a neighbourhood of $0$.

\section{Asymptotic behaviour of the test statistic}{\label{main}}

In order to derive the limit behaviour of the proposed test statistic,  we will assume that $w(t)$ has bounded support contained in some finite interval $\itI$  and we will introduce the empirical process given by $\itW_n=\left\{\itW_n(t,\ba)=\sqrt{n}P_n   \sin\left[ t\ba\trasp \bV_n^{-1/2} \left( \bX-\bm_n \right) \right]    \;, \;(t,\ba)\in \itI\times \itS_p\right\} $.
Theorem \ref{main}.1 states the asymptotic distribution of the process $\left\{W_n(t,\ba)\right\}$ under the null hypothesis, while Theorem \ref{main}.2 concerns the behaviour under local alternatives.   Note that in our statement we do not require neither finite moment conditions to the random vector $\bX$, nor a rate of convergence of the shape matrix estimator.  In this sense, our result provides an improvement over the proposal given in Zhu and Neuhaus (2003) who required finite fourth moment. As shown in Section \ref{monte}, the lack of moments may distort the results of the classical test based on $T_{n,\ml}$, while when using robust estimators the test is still reliable. On the other hand,  when second moments exist, if we take $\bm_n$ and $\bV_n$ as the sample mean and covariance matrix, respectively, Theorem \ref{main}.1 provides the asymptotic distribution of $W_n$ under slightly more general conditions than those given in Theorem 2.1 of Zhu and Neuhaus (2003).

\subsection{Behaviour under the null hypothesis}

\noi \textbf {Theorem \ref{main}.1.} \textsl{Let $\itI$ be a bounded interval. Assume that $\bX\sim P=\itE_p(\bmu, \bSi, \psi)$, i.e.,  that $H_0$ holds and that  $\int_0^1\sqrt{\log H_u}du <\infty$, where   $H_\delta$ is the smallest value $H\ge 1$ such that $\prob(\|\bSi^{-1/2}(\bX-\bmu)\|> H)\le \delta^2$, that is $H_\delta=\max(1,F_S^{-1}(1-\delta^2))$ with $F_S$  the distribution function of  $S=\|\bSi^{-1/2}(\bX-\bmu)\|$. Moreover, assume  that $\prob(\bX=\bmu)=0$ and that $\bV_n$ and $\bm_n$ are consistent estimators of $\bSi$ and $\bmu$, respectively, such that 
$\bm_n$ admits, for some function $\alpha_{\bm}:\real\to \real$,   a Bahadur expansion  as follows
\begin{equation}
\sqrt{n}(\bm_n- \bmu)= \frac 1{\sqrt{n}} \sum_{i=1}^n \left(\bX_i-\bmu \right) \;\alpha_{\bm} \left(\| \bSi^{-1/2}\left(\bX_i-\bmu \right)\|\right) +o_{\prob}(1)\;,
\label{vonmises}
\end{equation}
where $\esp_P \|\bSi^{-1/2}\left(\bX-\bmu \right) \|^2 \;\alpha_{\bm}^2  (\| \bSi^{-1/2}\left(\bX-\bmu \right)\|)<\infty$.
\newline Then,  the process $\itW_n=\left\{\itW_n(t,\ba)=\sqrt{n}P_n   \sin\left[ t\ba\trasp \bV_n^{-1/2} \left( \bX-\bm_n \right)   \right]  \;, \;(t,\ba)\in \itI\times \itS_p\right\} $ converges in distribution to a {centered} continuous Gaussian process $\itW=\{\itW(t,\ba)\;, \;(t,\ba)\in \itI\times \itS_p \} $ with covariance kernel given by $\esp_P[k(t,\ba,\bX) k(s,\bb,\bX) ]$, for $(t,\ba)\in \itI\times \itS_p$ and $(s,\bb)\in \itI\times \itS_p$, 
where $ k(t,\ba,\bx)=\sin \left[t \ba \trasp \bSi^{-1/2}\left(\bx-\bmu \right) \right] -\psi(t^2)\,t \,\ba \trasp \bSi^{-1/2} \left(\bx-\bmu \right) \;\alpha_{\bm} \left(\| \bSi^{-1/2}\left(\bx-\bmu \right)\|\right)$. }

\vskip0.1in
\noi \textbf {Remark \ref{main}.1.} Note that the classical location estimator, that is, the sample mean corresponds to $\alpha_{\bm}(t)=1$ and this is the situation considered in Theorem 2.1 of Zhu and Neuhaus (2003) which requires the existence of fourth moments. On the other hand, as shown in Hampel \textsl{et al.} (1986), if $\bm_n$ is an estimator related to a functional $\bm(P)$ that is affine equivariant there exists a real function $\alpha_{\bm}:\real^{+}\to \real$ such that its influence function equals
 $\IF\left(\bx_0,\bm,P_0\right)= (\bx_0-\bmu)\alpha_{\bm}(\|\bSi^{-1/2 }(\bx_0-\bmu)\|)$. In most cases, the influence function is bounded so  the assumption $\esp_P \|\bSi^{-1/2}\left(\bX-\bmu \right) \|^2 \;\alpha_{\bm}^2  (\| \bSi^{-1/2}\left(\bX-\bmu \right)\|)<\infty$ is satisfied and no moment conditions are required.
 Besides, as it is well known, under mild conditions, the influence function allows to obtain a Bahadur expansion for the location estimator (see Fernholz, 1983).
In particular, for the $S-$estimator (see Lopuha\"a, 1989), we have that $\alpha_{\bm}(t)=\beta^{-1}u_{\ese}(t)$ where $u_{\ese}(t)=\psi_{\ese}(t)/t$, $\psi_{\ese}(t)=\rho_{\ese}^{\prime}(t)$. Let $G_0$ be the spherical distribution related to $P_0$, that is, $G_0$ is the distribution of $\bSi^{-1/2 }(\bx_0-\bmu)$. Then, the constant $\beta$ is given by
\begin{eqnarray}
\beta &=&\esp_{G_0}\left[\left(1-\frac 1p\right)u_{\ese}(\|\bZ\|)+\frac 1p \psi_{\ese}^{\prime}(\|\bZ\|)\right]\;. \label{betas}
\end{eqnarray}
Usually, the influence function is computed at the central Gaussian distribution, so that $G_0=N({\bf{0}}, {\bf{I}})$. A common choice for the $\rho-$function defining the $S-$estimator is the Tukey function defined as $\rho_{\ese}(y)=(c^2/6)\min\{1-\left[1-(y/c)^2\right]^3, 1\}$.
Hence, $\psi_{\ese}(y)=y\left[1-\left(y^2/c^2\right)\right]^2 \mathbb{I}_{[-c,c]}(y)$,   $\psi_{\ese}^{\prime}(y)=\left[1-6\left(y^2/c^2\right)+5 \left(y^4/c^4\right)\right]\mathbb{I}_{[-c,c]}(y)$ and $u_{\ese}(y)=\left[1-\left(y^2/c^2\right)\right]^2 \mathbb{I}_{[-c,c]}(y)$ (see Lopuha\"a, 1989).

\vskip0.1in
\textbf {Remark \ref{main}.2.} Let us show that the assumption  $\int_0^1\sqrt{\log H_u}\,du <\infty$ where   $\prob(\|\bSi^{-1/2}(\bX-\bmu)\|> H_{\delta})\le \delta^2$, is fulfilled for  some distributions where fourth moments may not exist.  For the sake of simplicity, we will assume $\bmu=\bcero$ and $\bSi=\identidad_p$, since otherwise, we may consider $\bZ=\bSi^{-1/2}(\bX-\bmu)$.

 It is clear that if $\esp_{P}\|\bZ\|^2 <\infty $, then  $H_\delta \le   \left(\esp_{P}\|\bZ\|^2\right)^{1/2} /\delta$ and $\int_0^1\sqrt{\log H_u}du <\infty$. More generally, if $\esp_{P}\|\bZ\|^\nu <\infty $, for some $\nu>0$, then  $H_\delta \le  \left(\esp_{P}\|\bZ\|^\nu\right)^{1/\nu}/\delta^{1/\nu}$, so $\int_0^1\sqrt{\log H_u}\,du <\infty$. 

 As an example of elliptical distributions satisfying the condition  $\int_0^1\sqrt{\log H_u}\,du <\infty$, let us consider the multivariate $t-$distribution with $k$ degrees of freedom, i.e., $\bZ\sim \itT_{p,k}(\bcero,\identidad)$. As is well known $\bZ$  has no finite fourth moment when $k\le 4$. Besides, $\bZ$ has the same distribution as $y^{1/2} \bW $ where  $v_k=  k y^{-1}  \sim \chi^2_k$ and $\bW\sim N_p(\bcero, \identidad)$, where $\chi^2_k$  stands for the chi--square distribution with $k$ degrees of freedom.  Then, if the two expectations on the right hand side of (\ref{integral}) below exist, by the independence between $\bW$ and $y$ we have that
\begin{equation}
\esp  \|y^{1/2} \bW \|^\nu=k^{m}\esp\left( \frac{\| \bW \|^\nu}{v_k^{m}}\right)=k^{m}\;\esp \| \bW \|^\nu\, \esp\left(  {v_k^{-m}}\right)\,,
\label{integral}
\end{equation}
where $m=\nu/2$. Note that $\esp \| \bW \|^\nu<\infty$ for any $\nu>0$. On the other hand, using that  $v_k  \sim \chi^2_k$, we have that, for any $0<\nu<k$, $\esp(v_k^{-\nu/2})<\infty$, which entails that for the multivariate $t-$distribution   $\int_0^1\sqrt{\log H_u}\,du <\infty$. This result shows that our assumption is a very mild one since it includes for instance, the multivariate Cauchy distribution. 

\vskip0.1in
The following Corollary gives the distribution of the test statistic under the null hypothesis.

\vskip0.1in
\noi \textbf {Corollary \ref{main}.1.} \textsl{Under the assumptions of Theorem \ref{main}.1 if $w(t)$ is a weight function with bounded support $\itI$, then $T_n(\bm_n,\bV_n)\convdist \int_{\itS_p} \int_{\itI} \itW^2(t,\ba) w(t)\,dt\,dv(\ba) $ where the process $\itW(t,\ba)$ is defined in Theorem \ref{main}.1.}

\subsection{Behaviour under  the alternative}

 Regarding the consistency of the test, it is well known that if $\bX \sim \itE_p(\bmu, \bSi, \psi)$, then for any $\ba\in \itS_p$, $Z_{\ba}=\ba\trasp\bSi^{-1/2}(\bX-\bmu)$ has a symmetric distribution, but the converse is not true. A typical example being a random variable $\bY$ with distribution uniform on the set $\{\bx : -1\le x_j\le 1 \; \mbox{ for all } j\}$ which satisfies that $f_{\bY}(\by)=f_{\bY}(-\by)$ ensuring that all projections $\ba\trasp \bY$ are symmetric. 
In this situation, for observations having a symmetric distribution but not a spherical one, the test type--statistics considered in Zhu and Neuhaus (2003),  Ghosh and Ruymgaart (1992) or in this paper will not reject the null hypothesis. This is a feature  of any projection--pursuit procedure based on the property that any projection of a spherical distributed random vector is symmetric. 

On the other hand, as mentioned in Zhu and Neuhaus (2003), if $\esp_P \sin ( t \ba\trasp\bSi^{-1/2}(\bX-\bmu) ) \ne 0$, for some $t\in \itI$ and $\ba\in \itS_p$, then  $ T_{\bmuch,\bSich}(P)>0$. Therefore, using that $T_n(\bmu, \bSi)/n\convprob T_{\bmuch,\bSich}(P)$ together with the consistency of $\bm_n$ and $\bV_n$, we obtain that the test statistic $T_n(\bm_n,\bV_n)$ will converge to $\infty$ and the test is consistent against \textsl{global alternatives}.

\vskip0.1in
To derive the distribution of the test statistic under a set of alternatives, denote as $\sin^{(j)}(t)$ the $j-$th derivative of the sinus function at $t$. Recall that $\bV(P)$ and $\bm(P)$ stand for the functionals related to $\bV_n$ and $\bm_n$, respectively, when   $\bX\sim P$. We will assume that $\bm(P)$  is affine equivariant. To strength the dependence on the sample, we will denote as $P_{n,\bX}$ the empirical distribution of the sample $\bX_1,\dots, \bX_n$ and $\bV_{n,\bX}$ and $\bm_{n,\bX}$ the estimators based on that sample, that is,  $\bV_{n,\bX}=\bV(P_{n,\bX})$ and $\bm_{n,\bX}=\bm(P_{n,\bX})$.  Moreover, assume that the i.i.d. observation $\bX_i=\bX_{in}$ are such that $\bX_i=\bX_{in}=\bZ_i+\bY_i \,n^{-\alpha}$ for some $\alpha>0$ where $\bZ_i$ are i.i.d. such that $\bZ_i\sim P_0=\itE_p(\bmu, \bSi, \psi)$. Due to the equivariance of the location estimator and without loss of generality, we may assume that $\esp(\bY)=\bcero$. Effectively, if we define $\widetilde{\bX}_i=\bX_i-\esp(\bY)/n^\alpha$, using that  $\bm_{n,\widetilde{\bX}}=\bm_{n,\bX}-\esp(\bY)/n^\alpha$, we have that
$P_n\sin\left[t\ba\trasp\bV_n^{-1/2}(\widetilde{\bX}-\bm_{n,\widetilde{\bX}})\right]=P_n\sin\left[t\ba\trasp\bV_n^{-1/2}(\bX-\bm_{n,\bX})\right]$, so to obtain the asymptotic behaviour of the test statistic under these alternatives we may assume that  $\esp(\bY)=\bcero$.

\vskip0.1in
\noi \textbf {Theorem \ref{main}.2.} \textsl{Let   $w(t)$ be a weight function with  support $\itI$ and $\bZ\sim P_0=\itE_p(\bmu, \bSi, \psi)$. Assume that the following assumptions hold
\begin{enumerate}
\item[a)] $\bZ_i$ are i.i.d. such that $\bZ_i\sim \bZ$ and define $\bX_i=\bX_{in}=\bZ_i+\bY_i \,n^{-\alpha}$, where $\alpha>0$ and  $\esp(\bY)=\bcero$. 
\item[b)]  $\prob(\bZ=\bmu)=\prob(\bX=\bmu)=0$, 
\item[c)]  $\int_0^1\sqrt{\log H_u}du <\infty$, where   $H_\delta$ is the smallest value $H\ge 1$ such that $\prob(\|\bSi^{-1/2}(\bZ-\bmu)\|> H)\le \delta^2$, that is $H_\delta=\max(1,F_S^{-1}(1-\delta^2))$ where $F_S$ is the distribution function of  $S=\|\bSi^{-1/2}(\bZ-\bmu)\|$. 
\item[d)]  $\bV_{n,\bX}$ and $\bm_{n,\bX}$ are such that $\bV_{n,\bX}\convprob \bSi$ and $\bm_{n,\bX}- \bmu\convprob 0$.
\item[e)] The functional $\bm(P)$ is such that the  Bahadur expansion  (\ref{vonmises}) holds at $P_0$, that is, for some function $\alpha_{\bm}:\real\to \real$,   $\bm_{n,\bZ}$ satisfies (\ref{vonmises}) where    $\esp_{P_0} \|\bSi^{-1/2}\left(\bZ-\bmu \right) \|^2 \;\alpha_{\bm}^2  (\| \bSi^{-1/2}\left(\bZ-\bmu \right)\|)<\infty$. Furthermore,   
\begin{equation}
 \sqrt{n}(\bm_{n,\bX}- \bmu ) = \sqrt{n}(\bm_{n,\bZ}- \bmu)+ o_{\prob}(1)\;.   
\label{condicion_mux}
\end{equation}
\item[f)] there is  {a positive} integer $m$ such that $\esp \|\bY\|^{2m}<\infty $  and for some $\delta_m>0$
\begin{eqnarray*}
\sup_{(t,\ba)\in \itI^{\delta_m}\times \itS_p} |B_m(t,\ba)|&=&\sup_{(t,\ba)\in \itI^{\delta_m}\times \itS_p}\left| \esp\left\{ \left(t\ba\trasp \bSi^{-1/2}\bY\right)^m \sin^{(m)}(t\ba\trasp\bSi^{-1/2}(\bZ-\bmu))\right\}\right|\ne 0\,,
\end{eqnarray*}
where $\itI^{\delta}$ stands for a $\delta-$neighborhood of $\itI=[-\nu,\nu]$.
\end{enumerate}
Let $\ell$ be the smallest {positive} $m$ satisfying f). If $\bY$ and $\bZ$ are not independent assume in addition that for any $1\le s<\ell$,   $\int_0^{1} \sqrt{\log H_{u^{q}}}du<\infty$, where $q= {\ell}/({\ell-s})$.  Then, when $\alpha=1/(2\,\ell)$,    
$$T_n(\bm_n,\bV_n)\convdist \int_{\itS_p} \int_{\itI} \left(\itW(t,\ba)+\frac{1}{\ell!}B_\ell(t,\ba)\right)^2 w(t)\,dt\,dv(\ba) \,,$$
 where the process $\itW(t,\ba)$ is defined in Theorem \ref{main}.1.}

 
\section{Bootstrap method}{\label{boot}}
As mentioned in Zhu and Neuhaus (2003), the asymptotic behaviour of the test statistic does not allow to compute easily $p-$values, so a bootstrap method is needed. Zhu and Neuhaus (2003) describe a bootstrap procedure when the center $\bmu$ and the shape parameter $\bSi$ are known and when $\bmu$ is estimated using the mean of the observations. When the center  and the shape matrix are unknown, a slight modification to the method considered in Zhu and Neuhaus (2003)  is needed to adapt to the resistant location estimators, since the estimated shape parameter does not influence the distribution of the test statistic.

One possibility is to adapt  the bootstrap statistic defined in Zhu and Nehaus (2003) to the present setting. For that purpose, assume, as in Section \ref{main}, that the affine equivariant location estimator admits  a Bahadur expansion given by (\ref{vonmises}). Let $\wbd= \ba\trasp \bV_n^{-1/2}$. Using that 
\begin{eqnarray*}
P_n    \sin \left[t \wbd \trasp\left( \bX-\bm_n \right)\right] & = &\cos\left[t \wbd \trasp \left( \bmu-\bm_n \right)\right]P_n  \sin \left[t \wbd \trasp\left( \bX-\bmu \right)\right] \\
&& -\sin\left[t \wbd \trasp \left( \bm_n-\bmu \right)\right]P_n \cos \left[t \wbd \trasp\left( \bX-\bmu \right)\right] \; ,
\end{eqnarray*}
the first order von Mises expansion (\ref{vonmises})
 and the fact that $ \bm_n   \convprob \bmu$, we have that
$$
\sqrt{n}P_n   \sin \left[t \wbd \trasp\left( \bX-\bm_n \right)\right]  =  \sqrt{n} P_n  \sin \left( t \ba\trasp \bz \right)- \sqrt{n} \sin \left \lbrace  t \ba\trasp  P_n \left[ \bz \alpha_\bm \left( \| \bz\|\right)\right]\right\rbrace  P_n \left[ \cos \left(t \ba\trasp \bz \right)\right] +o_{\prob}(1),
$$
where $\bz=\bSi^{-1/2}\left(\bX-\bmu \right)\sim \bu \| \bSi^{-1/2} \left( \bX-\bmu \right)\|$ and $\bu \sim \itU(\itS_p)$.

However, the implementation of this bootstrapping method when using a location and scatter matrix robust estimators  implies the computation of $\alpha_{\bm}(t)$, which equals $1$ for the sample mean, but may be more complex when using robust estimators. For instance, when using $S-$estimators the function $\alpha_{\bm}(t)$ involves the calculation of the constant $\beta$ defined in  (\ref{betas}). In robustness, $\beta$ is usually  computed under the standard normal distribution. However, in our situation, the constant $\beta$ must be computed under the spherical distribution related to the underlying elliptical distribution of the sample. This is a drawback of this bootstrap method since the correct distribution is unknown. To avoid this vicious circle, we consider a bootstrap statistic that can be computed as follows:

\begin{enumerate}
\item[\textbf{Step 1}] Generate i.i.d. random vector $\bu_i \sim \itU(\itS_p)$ and define $\bU_n=(\bu_1,\dots,\bu_n)$. Let \linebreak $\bX_i^{\star}= \bu_i \| \bV_n^{-1/2} \left( \bX_i-\bm_n \right)\|$ be the bootstrap observations  and $P_n^{\star}$ the empirical distribution of $\bX_i^{\star}$
\item[\textbf{Step 2}] Define $W_n^{\star}(t,\ba)= \sqrt{n}P_n^{\star}  \sin\left[ t\ba\trasp {{\bV}^*}^{-1/2} \left( \bX^{\star}-\bm^* \right) \right]  $,
where $\bm^*$ and $\bV^*$ are the location and scatter matrix robust estimators of the bootstrapped sample $\bX_i^{\star}$ and calculate the statistic $
T_{n, \bm^{\star}, \bV^{\star}}^{\star}(\bU_n)=\int_{\itS_p} \int W_n^{\star}(t,\ba)^2\;w(t)dtdv(\ba)$.
\item[\textbf{Step 3}] Repeat Steps 1 and 2 $Nboot$ times to get $Nboot$ values of $
T_{n,\bm^{\star}, \bV^{\star}}^{\star}(\bU_n^{(j)})$, $1\le j\le Nboot$.
\item[\textbf{Step 4}] Estimate the $p-$value as $p=k/(Nboot+1)$ where $k$ is the number of $T_{n,\bm^{\star}, \bV^{\star}}^{\star}(\bU_n^{(j)})$ that are greater or equal than $T_{n, \bm,\bV }$.
\end{enumerate}

Through this algorithm we obtain a sample of  bootstrap replicates $T^*_j$, $1 \le j \le Nboot$ 
whose distribution approximates the distribution of $T_{n, \bm,\bV }$ under the null hypothesis, as desired.

The proof of the asymptotic distribution of the boostrap procedure is an interesting topic which we leave for future research.

\section{Finite sample distribution of the test statistic when $p=2$}{\label{numerical}}

We  generate   independent observations $\bZ_1, \dots ,\bZ_n$, $\bZ_i \sim P$ according to different elliptical models under the null hypothesis. Let ${\cal T}_{p,k}(\bmu, \bSi)$ be the multivariate $p-$dimensional $t-$distribution with $k$ degrees of freedom, which includes the multivariate Cauchy distribution when $k=1$,  and denote $\itU(\itS_p) $ and $\itU(\itB_p)$  the uniform distributions over the unit circle and the unit ball, respectively. Consider  the null hypotheses
$H_0^{(1)}:$ $P = N_p({\bf{0}},\bI)$,
$H_0^{(2)}:$ $P = 0.9 N_p({\bf{0}},\bI) +0.1 {\cal T}_{p,1}({\bf{0}}, \bI)$,   
$H_0^{(3)}:$ $P = 0.9 N_p({\bf{0}},\bI) +0.1 {\cal T}_{p,3}({\bf{0}}, \bI)$, $H_0^{(4)}:$ $P = {\cal T}_{p,3}({\bf{0}}, \bI)$, $H_0^{(5)}:$ $P = \itU(\itS_p) $,  $H_0^{(6)}:$ $P = \itU(\itB_p)$ 
and $H_0^{(7)}:$ $P = {\cal T}_{p,1}({\bf{0}}, \bI)$. 

For each null hypothesis $H_0^{(j)}$, we consider different  alternative hypothesis $H_{1,\Delta}^{(j)}$, related to the original distribution $P$ in the null hypothesis. Under  $H_{1,\Delta}^{(j)}$, the observations are generated as $\bX_i=\bZ_i+\Delta \bY_i$ with $\bZ_i \sim P$ independent of $\bY_i$ and $\bY=(Y_1,Y_2)\trasp$ where $Y_k\sim \chi^2_1$ independent among each other and $\Delta= 0.5$, $1$ and $1.5$. We also studied the behaviour of the statistics under  two fixed alternatives  $H_1^{\star^{(1)}}$ and  $H_1^{\star^{(2)}}$. Under $H_1^{\star^{(1)}}$, the data have the   distribution   of a random vector with two independent components, $\itE(1)$ and $N(0,1)$, where $\itE(\lambda)$ denotes the exponential distribution of parameter $\lambda$, that is, with mean value $1/\lambda$, while the alternative  $H_1^{\star^{(2)}}$  corresponds to  the distribution of a random vector with two independent components, $\itE(1)$ and $\itE(1/2)$, that is, with expectation 1 and 2, respectively.
The first three alternatives were studied in Zhu and Neuhaus (2003), while Koltchinskii and Li (1998) studied the capability of their proposal to detect  $H_1^{\star^{(1)}}$  and  $H_1^{\star^{(2)}}$.

In all cases, we perform  $N=1000$ replications for samples of size $n=20$, $50$, $100$ and $200$. For each sample, we compute the test statistics with the mean and sample covariance matrix, denoted by $T_{n,\ml}$, with  the Donoho--Stahel estimators of location and scatter, denoted by $T_{n,\ds}$ and with the  $S-$estimators of location and scatter, denoted by $T_{n,\ese}$. Both robust estimators are calibrated to attain 50\% breakdown point. We choose as weight function $w(t)=\indica_{[-b,b]}(t)$ with $b=2$.

In Figures \ref{fig:densidadC0} to \ref{fig:densidadC2}, the density estimates of test statistics $T_{n,\ml}$, $T_{n,\ds}$ and $T_{n,\ese}$  are plotted under the null hypotheses $H_0^{(1)}$ to $H_0^{(7)}$ and under their corresponding alternatives. The density estimates were evaluated using the normal kernel.

 As expected, in most cases the classical test statistics is more sensitive to the lack of elliptical symmetry of the alternative distributions than the robust test statistics.
However, for $n=20$ in the considered situations all statistics, the classical and the robust ones, fail to distinguish the symmetric distribution under the null hypothesis from those considered in the alternatives.  Indeed, for this sample size all the density estimates are almost overlapping. The tests  detect some of the selected alternatives for $n=50$. For $n=100$, in all cases the ability of the test statistics to make out the nature of the underlying distribution  increases and this fact becomes more clear for $n=200$. For $n=100$ and $200$, the densities corresponding to the non--elliptical distributions generated under $H_1^{\star^{(1)}}$ and  $H_1^{\star^{(2)}}$  are shifted to the right from those of  the test statistic under the null hypothesis. This effect is less visible for $n=50$. Hence, one could expect that the tests statistics will work well under these circumstances. On the opposite, except for Figure \ref{fig:densidadC6},  the densities of all the test statistics under the null hypothesis and under the alternative $H_{1, 0.5}^{(j)}$ are almost overlapping. As expected, this performance is even worst for the classical test under $H_0^{(7)}$ and $H_{1,\Delta}^{(7)}$, where the distribution of the test statistic does not allow to distinguish between the null and the alternative hypotheses even for $n=200$ (see Figure \ref{fig:densidadC2}). Hence, one can not expect a good performance of the classical tests in this case. A similar conclusion can be held for  $H_{1, 1}^{(j)}$ for $n=100$, while for $n=200$ the behaviour of the test statistic depends on the distribution of $\bZ_i$. 
 
This numerical approach suggests that, for $p=2$,  small sample sizes and  values of $\Delta$ smaller than 0.5 when considering alternatives of the form $\bZ_i+\Delta \bY_i$  should not be considered in the Monte Carlo study presented in Section \ref{monte}.

\section{Monte Carlo study}\label{monte}
In this section, we report the results of a simulation study conducted to analyse the performance of the  test statistic obtained using robust location and scatter estimators with respect  to that   based on the sample mean and covariance matrix. The weight function considered in this Monte Carlo study equals $w(t)=\indica_{[-b,b]}(t)$ where $b=2$. Based on the results reported in Section \ref{numerical} regarding the finite--sample distribution of the test statistics, we carried out $NR=500$ replications  for sample sizes equal to $n=200$.   To perform the bootstrap method described in Section \ref{boot}, we used $Nboot=1000$ bootstrap samples. The nominal level was set equal to $\alpha=0.05$. Besides, we also compare our test procedure with other known methods for testing elliptical symmetry, when $p=5$. {In what follows, $\pi_{H_0}(T_n)$ and $\pi_{H_1}(T_n)$ stand for the observed probabilities of rejection of the test based on the statistic $T_n$ under the null hypothesis and under the alternative $H_1$, respectively. As in Batsidis \textsl{et al.} (2014), we also examine if the empirical size  is significantly different from the nominal level $\alpha=0.05$. To be more precise, let $\pi$ be such that $\pi_{H_0}(T_n)\convprob \pi$. Then, using the central limit theorem,   the hypothesis  $H_{0,\pi}: \pi=\alpha$ is rejected at level $\gamma$ versus  $H_{1,\pi}:\pi\ne\alpha$  if $\pi_{H_0}(T_n)\notin [a_1(\alpha),a_2(\alpha)]$ where $a_j(\alpha)=\alpha+ (-1)^j z_{\gamma/2}\,\{\alpha(1-\alpha)/NR\}^{1/2}$, $j=1,2$. If $H_{0,\pi}: \pi=\alpha=0.05$  is not rejected, the testing procedure based on $T_n$ is considered accurate. Note that if $\pi_{H_0}(T_n)<a_1(\alpha)$  the testing procedure is  conservative, while if $\pi_{H_0}(T_n)>a_2(\alpha)$ the test is liberal. In  all Tables reporting the observed frequencies of rejection, we indicate with $\star$ those cases  in which the observed empirical frequencies of rejection are   different from the nominal level with a significance level $\gamma=0.01$.}

From now on, let ${\cal T}_{p,k}(\bmu, \bSi)$ denote the multivariate $p-$dimensional $t-$distribution with $k$ degrees of freedom, which includes the multivariate Cauchy distribution when $k=1$,  and denote $\itU(\itS_p) $ and $\itU(\itB_p)$  the uniform distributions over the unit circle and the unit ball, respectively. Denote also as $\chi_\nu^2$ the chi--square distribution with $\nu$ degrees of freedom.

\subsection{Simulation study in dimension $p=2$}{\label{montecarlo}}
We  generate   independent observations $\bZ_1, \dots ,\bZ_n$, $\bZ_i \sim P$ according to different elliptical models under the null hypothesis.   Consider  the null hypotheses
$H_0^{(1)}:$ $P = N_p({\bf{0}},\bI)$,
$H_0^{(2)}:$ $P = 0.9 N_p({\bf{0}},\bI) +0.1 {\cal T}_{p,1}({\bf{0}}, \bI)$,   
$H_0^{(3)}:$ $P = 0.9 N_p({\bf{0}},\bI) +0.1 {\cal T}_{p,3}({\bf{0}}, \bI)$, $H_0^{(4)}:$ $P = {\cal T}_{p,3}({\bf{0}}, \bI)$, $H_0^{(5)}:$ $P = \itU(\itS_p) $,  $H_0^{(6)}:$ $P = \itU(\itB_p)$ 
and $H_0^{(7)}:$ $P = {\cal T}_{p,1}({\bf{0}}, \bI)$. 

For each null hypothesis $H_0^{(j)}$, we consider different  alternative hypothesis $H_{1,\Delta}^{(j)}$, related to the original distribution $P$ in the null hypothesis. Under  $H_{1,\Delta}^{(j)}$, the observations are generated as $\bX_i=\bZ_i+\Delta \bY_i$ with $\bZ_i \sim P$ independent of $\bY_i$ and $\bY=(Y_1,Y_2)\trasp$ where $Y_k\sim \chi^2_1$ independent among each other and $\Delta= 0.5$, $1$ and $1.5$. We also studied the behaviour of the statistics under  two fixed alternatives  $H_1^{\star^{(1)}}$ and  $H_1^{\star^{(2)}}$. Under $H_1^{\star^{(1)}}$, the data have the   distribution   of a random vector with two independent components, $\itE(1)$ and $N(0,1)$, where $\itE(\lambda)$ denotes the exponential distribution of parameter $\lambda$, that is, with mean value $1/\lambda$, while the alternative  $H_1^{\star^{(2)}}$  corresponds to  the distribution of a random vector with two independent components, $\itE(1)$ and $\itE(1/2)$, that is, with expectations 1 and 2, respectively.
The first three alternatives were studied in Zhu and Neuhaus (2003), while Koltchinskii and Li (1998) studied the capability of their proposal to detect  $H_1^{\star^{(1)}}$  and  $H_1^{\star^{(2)}}$.
 
For each sample, we compute the $p-$values of the test statistics obtained using  the mean and sample covariance matrix, denoted by $T_{n,\ml}$,    the Donoho--Stahel estimators of location and scatter, denoted by $T_{n,\ds}$ and  the  $S-$estimators of location and scatter, denoted by $T_{n,\ese}$. Both robust estimators are calibrated to attain 50\% breakdown point.  The corresponding frequencies of rejection are reported in Tables \ref{tab:nivel_ml} and \ref{tab:nivel_ds}, where $\Delta=0$ corresponds to the observations generated according to the null hypothesis.   

Taking as reference the first row of Table \ref{tab:nivel_ml}, as expected, we observe some loss of power of the classical test based on $T_{n,\ml}$ under the alternatives of the distributions considered in $H_0^{(2)}$ and $H_0^{(4)}$,  where the data follow heavier tailed distributions. On the other hand, the opposite is observed when considering $\itU(\itS_p)$ and $\itU(\itB_p)$. The extreme situation is found under $H_0^{(7)}$ and  its alternatives, since the classical test completely looses its power. Indeed, in this case this test is unable to distinguish between data coming from the elliptical distribution ${\cal T}_{2,1}({\bf{0}}, \bI)$ from data generated  under its alternatives $H_{1,\Delta}^{(7)}$ for $\Delta=0.5, 1$ and $1.5$.  Besides, Table \ref{tab:nivel_ds} shows that for both families of robust estimators of location and scatter matrix similar results are obtained, either  in level or power. Indeed, with both robust estimators, the proposed tests lead to comparable results to those obtained with the classical test for $H_0^{(j)}$ for $j=1,3, 5$ and $6$, even when for $j=1$ and $3$ there is some loss of power under $H_{1,0.5}^{(j)}$. However, the robust tests outperform the behaviour observed with $T_{n,\ml}$ under the alternatives of $H_0^{(2)}$ and $H_0^{(4)}$, getting larger frequencies of rejection. Finally, from Table \ref{tab:nivel_ds} we conclude that using $T_{n,\ds}$ and $T_{n,\ese}$ the decision rule has a good performance and is   informative even under $H_0^{(7)}$ and  its alternative hypotheses. 

 We have also considered two other alternatives also studied in Batsidis \textsl{et al.} (2014) which gave power 1 as  $H_1^{\star^{(1)}}$ and $H_1^{\star^{(2)}}$, for that reason the results are omitted in the Tables. One of the alternatives, denoted as $H_1^{\star^{(3)}}$,  is obtained generating random vectors with  two independent components with a common beta distribution $Be(5,1)$. The other one, $H_1^{\star^{(4)}}$, corresponds to  the distribution of a random vector with distribution  $0.5 \, N_p({\bf{0}},\bI) +0.5\, N_p(\bmu, \bSi)$ with $\bmu=(1,2)$ and $\bSi=\left(\begin{array}{cc}5 & -4 \\ -4 & 5 \end{array}\right)$.

\footnotesize
\begin{table}[ht!]
\begin{center}\footnotesize
\begin{tabular}{  c c c c c c c}
  & \multicolumn{6}{c}{$T_{n,\ml}$} \\\hline 
  & & \multicolumn{3}{c}{$H_{1,\Delta}^{(j)}$}& $H_1^{\star^{(1)}}$ & $H_1^{\star^{(2)}}$ \\\cline{3-5}
$\Delta$& $ 0$ & $ 0.5$ & $ 1$ & $ 1.5$  & & \\\hline
$H_0^{(1)}$  & 0.060 & 0.478 & 1.000  &  1.000  & 1.000 & 1.000\\ 
$H_0^{(2)}$& 0.086  & 0.220  & 0.748  &  0.888 & 1.000 & 1.000\\   
$H_0^{(3)}$ & 0.048 & 0.412 & 0.994 & 0.998  & 1.000 & 1.000\\ 
$H_0^{(4)}$ &  0.044 & 0.110 & 0.794 & 0.988  & 1.000 & 1.000\\ 
$H_0^{(5)}$ & 0.057 & 0.892 & 1.000 & 1.000   & 1.000 & 1.000\\ 
$H_0^{(6)}$ & 0.054 & 1.000 & 1.000 & 1.000   & 1.000 & 1.000\\ 
$H_0^{(7)}$ &  0.056 & 0.058 & 0.062 & 0.076   & 1.000 & 1.000\\ 
\hline
\end{tabular}
\end{center}
\caption{\label{tab:nivel_ml} \footnotesize Frequency of rejection for the bootstrap test $T_{n,\ml}$ for $n=200$ and dimension $p=2$. {$\star$ indicates that the frequency of rejection is significantly different from the nominal level.}}
\end{table}
\normalsize

\footnotesize
\begin{table}[ht!]
\begin{center}\footnotesize
\begin{tabular}{  c c c c c c cc c c c c c c}
&\multicolumn{6}{c }{$T_{n,\ds}$}&&\multicolumn{6}{c}{$T_{n,\ese}$}\\\hline
  & & \multicolumn{3}{c}{$H_{1,\Delta}^{(j)}$}& $H_1^{\star^{(1)}}$ & $H_1^{\star^{(2)}}$ & & &\multicolumn{3}{c}{$H_{1,\Delta}^{(j)}$}& $H_1^{\star^{(1)}}$ & $H_1^{\star^{(2)}}$\\\cline{3-5}\cline{9-11}
$\Delta$& $ 0$ & $ 0.5$ & $ 1$ & $ 1.5$ & & & &$ 0$ & $ 0.5$ & $ 1$ & $1.5$ & &\\\hline
$H_0^{(1)}$  & 0.048 & 0.256 &  0.990  & 1.000  & 1.000 & 1.000& &  0.060 & 0.270& 0.988 &  1.000  &  1.000 & 1.000\\
$H_0^{(2)}$  & 0.058  & 0.283  &  0.984 &  1.000 & 1.000 & 1.000& & 0.060  & 0.309  & 0.980  & 1.000 & 1.000 & 1.000\\ 
$H_0^{(3)}$ & 0.042 & 0.244 & 0.984 & 1.000   & 1.000 & 1.000& & 0.050 & 0.264 & 0.986 & 1.000  & 1.000 & 1.000\\ 
$H_0^{(4)}$  & 0.056 & 0.206 & 0.862 & 0.998  & 1.000 & 1.000 & & 0.056 & 0.212 & 0.876 & 0.998 & 1.000 & 1.000\\ 
$H_0^{(5)}$  & 0.062 & 0.608 & 1.000 & 1.000 & 1.000 & 1.000&  &0.068 & 0.552 & 1.000 & 1.000 & 1.000 & 1.000\\ 
$H_0^{(6)}$  & 0.048 & 1.000 & 1.000 & 1.000  & 1.000 & 1.000& & 0.050 & 1.000 & 1.000 & 1.000   & 1.000 & 1.000\\ 
$H_0^{(7)}$ &  0.048 & 0.066 &  0.464 & 0.832    & 1.000 & 1.000 &   & 0.046 & 0.068 & 0.440 & 0.840   & 1.000 & 1.000\\ 
\hline
\end{tabular}
\end{center}
\caption{\label{tab:nivel_ds} \footnotesize Frequency of rejection for the bootstrap test  $T_{n,\ds}$ and $T_{n,\ese}$ for $n=200$ and dimension $p=2$.  {$\star$ indicates that the frequency of rejection is significantly different from the nominal level.}}
\end{table}
\normalsize

 As noted before,  the exact sizes of the test statistics, i.e., $\pi_{H_0}(T_{n,\ml})$, $\pi_{H_0}(T_{n,\ds})$ and $\pi_{H_0}(T_{n,\ese})$ fluctuate around the fixed level $\alpha=0.05$.
 To help in the visual comparison of the power performance of the three test  statistics, as in Batsidis \textsl{et al.} (2014), Table \ref{tab:rho_dim2} reports the size--corrected relative exact powers $\rho_{H_1}(T_{n,\ds},T_{n,\ml})$ and $\rho_{H_1}(T_{n,\ese},T_{n,\ml})$. For two test statistics, $T_n^{(1)}$ and $T_n^{(2)}$,  $\rho_{H_1}(T_n^{(1)},T_n^{(2)})$ was defined in Morales \textsl{et al.} (2004) as
\begin{equation}
\rho_{H_1}(T_n^{(1)},T_n^{(2)})=\left(\frac{D_{H_1}(T_n^{(1)})}{D_{H_1}(T_n^{(2)})}-1\right)\times 100\,,
\label{rho}
\end{equation}
with $D_{H_1}(T_{n})=\pi_{H_1}(T_n)-\pi_{H_0}(T_n)$. This measure allows to clarify the fluctuations in the powers which are more difficult to observe in Tables \ref{tab:nivel_ml} and \ref{tab:nivel_ds}.
\footnotesize 
\begin{table}[ht!]
\begin{center}\footnotesize
\begin{tabular}{  c|  r r r r r|  r r r r r  }
  & \multicolumn{5}{c}{$\rho_{H_1}(T_{n,\ds},T_{n,\ml})$} & \multicolumn{5}{c}{$\rho_{H_1}(T_{n,\ese},T_{n,\ml})$}\\\hline 
  &   \multicolumn{3}{c}{$H_{1,\Delta}^{(j)}$}& $H_1^{\star^{(1)}}$ & $H_1^{\star^{(2)}}$  &   \multicolumn{3}{c}{$H_{1,\Delta}^{(j)}$}& $H_1^{\star^{(1)}}$ & $H_1^{\star^{(2)}}$ \\\cline{2-4}\cline{7-9}
$\Delta$&  \multicolumn{1}{c}{$ 0.5$} & \multicolumn{1}{c}{$ 1$} & \multicolumn{1}{c}{$ 1.5$}  & & &  \multicolumn{1}{c}{$ 0.5$} & \multicolumn{1}{c}{$ 1$} & \multicolumn{1}{c}{$ 1.5$}  & &\\\hline
$H_0^{(1)}$  & -50.239  &  0.213 &  1.277   & 1.277  & 1.277 &-49.761 & -1.277 &  0.000 &  0.000  &  0.000   \\\ 
$H_0^{(2)}$ & 67.910  & 39.879  & 17.456  & 3.064 &  3.064 & 85.821 & 38.973 &17.207 & 2.845 & 2.845\\   
$H_0^{(3)}$ & -44.506 & -0.423 &  0.842   & 0.630  &  0.630  & -41.209 & -1.057  &  0.000  & -0.210 &  -0.210 \\ 
$H_0^{(4)}$ & 127.273 &  7.467 &  -0.212 & -1.255  & -1.255 & 136.364 &  9.333  & -0.212 & -1.255  & -1.255  \\ 
$H_0^{(5)}$ & -34.611 &  -0.530  & -0.530  & -0.530  & -0.530  & -42.036 & -1.167 & -1.167 & -1.167 &  -1.167\\ 
$H_0^{(6)}$  & 0.634  & 0.634  & 0.634  & 0.634  & 0.634  &0.423& 0.423 & 0.423 & 0.423 & 0.423\\ 
$H_0^{(7)}$  &  800.000  & 6833.333  & 3820.000  &  0.848 &   0.848 & 1000.000 & 6466.667 & 3870.000   & 1.059     & 1.059 \\ 
\hline
\end{tabular}
\end{center}
\caption{\label{tab:rho_dim2} \footnotesize Size corrected relative exact power for the robust bootstrap tests $T_{n,\ds}$ and $T_{n,\ese}$ with respect to the classical one $T_{n,\ml}$ for $n=200$ and dimension $p=2$.}
\end{table}
\normalsize

 Table \ref{tab:rho_dim2} shows that in most cases, larger values of $\rho_{H_1}(T_{n},T_{n,\ml})$ are obtained with the Donoho--Stahel estimators over the $S-$estimators leading to the conclusion that the test based on the  Donoho--Stahel estimators is a preferable choice. As noted before, the robust tests outperform the classical one specially for alternatives close to the null hypothesis under $H_0^{(2)}$, $H_0^{(4)}$, $H_0^{(6)}$ and $H_0^{(7)}$. This performance was expected for the heavy tailed distributions $H_0^{(2)}$ and $H_0^{(7)}$, but it is also present under moderate tails as those of the ${\cal T}_{p,3}(\mbox{\bf 0}, \identidad)$ since fourth moments  do not exist. On the other hand,  as expected, the classical test has a superior behaviour under a Gaussian distribution,   when $\Delta=0.5$. The better performance of   $T_{n,\ml}$ for $\Delta=0.5$, is also observed under  $H_0^{(3)}$ and the uniform distribution over the unit circle. With respect to the detection of the alternatives $H_1^{\star^{(1)}}$ and $H_1^{\star^{(2)}}$, all procedures are almost equivalent. 
 Based on the simulated results obtained for the considered distributions, we recommend  the test statistic  based on the  Donoho--Stahel estimators. For this reason, in dimension $p=5$ we only compare  the test statistics $ T_{n,\ds}$ and $T_{n,\ml}$.

\subsection{Simulation study in dimension $p=5$}{\label{montecarlodim5}}
In order to compare the performance of the two test statistics  $T_{n,\ml}$ and $T_{n,\ds}$, under the null hypothesis, we  generate $n$ independent observations $\bZ_1, \dots ,\bZ_n$, $\bZ_i\sim P$, $\bZ_i\in \real^5$,  following  different elliptical distributions   as follows  $H_0^{(1)}:$ $P = N_p({\bf{0}},\bI)$,
  $H_0^{(2)}:$ $P$ is the  Pearson type II distribution  generated as $\sqrt{V} \bU$ where $\bU \sim \itU(\itS_p) $ and $V\sim Be( p/2,m)$,   with $m=3/2$,
 $H_0^{(3)}:$   $P = {\cal T}_{p,5}({\bf{0}}, \bI)$ and
$H_0^{(4)}:$   $P = {\cal T}_{p,1}({\bf{0}}, \bI)$.
As in   Section \ref{montecarlo}, we consider observations $\bX_i$, $i=1,\dots, n$ generated under  the alternative hypotheses $H_{1,\Delta}^{(j)}$, with $\Delta={0.25}, 0.5, {0.75}, 1$ and $1.5$. Besides, we studied the performance under four fixed alternatives $ H_1^{\star^{(j)}}$ for $j=1$ to $4$ defined  as follows.  Under $H_1^{\star^{(1)}}$, the data have the   distribution   of a random vector with $p$ independent components, the first $p-1$ having distribution $\itE(1)$ and the last one $N(0,1)$. This distribution corresponds to $H_1^{\star^{(1)}}$  in dimension $p=2$.  
 The second fixed alternative $H_1^{\star^{(2)}}$  corresponds to  the distribution of a random vector $\bX$ with $p$ independent components each of them with distribution  $\itE(1)$.  
 Under $H_1^{\star^{(3)}}$, $\bX_i\sim \bX$ where $\bX$ is a random vector with $p$ independent components with common distribution $Be(5,1)$.  
 Finally, $H_1^{\star^{(4)}}$ corresponds to the situation in which $\bX$ has  $p$ independent components, the first $p-1$ with common distribution $\itE(1)$ and the last one ${\cal T}_{p,1}$.  
 
The frequencies of rejection are reported in Table \ref{tab:nivel_mldsp5005}, where $\Delta=0$ corresponds to the observations generated according to the null hypothesis. Besides, Table \ref{tab:rho_mldsp5005} reports the size--corrected relative exact powers $\rho_{H_1}(T_{n,\ds},T_{n,\ml})$ as defined in (\ref{rho}).

\footnotesize 
\begin{table}[ht!]
\begin{center}\footnotesize
\begin{tabular}{  c| c cc c c c c c c c}
  & \multicolumn{10}{c}{$T_{n,\ml}$} \\\hline 
  & & \multicolumn{5}{c}{$H_{1,\Delta}^{(j)}$} & $H_1^{\star^{(1)}}$  & $H_1^{\star^{(2)}}$   & $H_1^{\star^{(3)}}$   & $H_1^{\star^{(4)}}$ \\\cline{3-7}
$\Delta$& $ 0$ & $0.25$ & $ 0.5$ & $0.75$ & $ 1$ & $ 1.5$  & & & &\\\hline
$H_0^{(1)}$  & 0.060 & {0.088} & 0.646 &{0.998} & 1.000  & 1.000 & 1.000 & 1.000 & 1.000&1.000   \\ 
$H_0^{(2)}$&  0.046 & {0.996} & 1.000 & {1.000} & 1.000  & 1.000 & 1.000 & 1.000 &1.000 &   1.000   \\   
$H_0^{(3)}$ & 0.080{$\star$} & {0.070} & 0.290 & {0.866} &0.996  &1.000 & 1.000 & 1.000 & 1.000& 1.000  \\ 
$H_0^{(4)}$ &  0.058 & {0.078} & 0.080  &{0.086} &  0.094 & 0.118 &1.000 & 1.000 &1.000 &1.000  \\ 
\hline
  & \multicolumn{10}{c}{$T_{n,\ds}$} \\\hline 
  & & \multicolumn{5}{c}{$H_{1,\Delta}^{(j)}$} & $H_1^{\star^{(1)}}$  & $H_1^{\star^{(2)}}$   & $H_1^{\star^{(3)}}$  & $H_1^{\star^{(4)}}$  \\\cline{3-7}
$\Delta$& $ 0$ & $0.25$ & $ 0.5$ & $0.75$ & $ 1$ & $ 1.5$  & & & &\\\hline
 $H_0^{(1)}$  & 0.034& {0.044} & 0.428 &{0.984} &  1.000 & 1.000 & 1.000& 1.000 &1.000 & 1.000   \\ 
$H_0^{(2)}$& 0.036 & {0.982} & 1.000 & {1.000} &1.000  & 1.000& 1.000 & 1.000  & 1.000& 1.000 \\   
$H_0^{(3)}$ & 0.040 & {0.052} & 0.270  & {0.870} &0.996  & 1.000& 1.000 & 1.000 &1.000 &1.000   \\ 
$H_0^{(4)}$ &  0.054 & {0.056} & 0.126 & {0.398} & 0.766  &  0.994 & 1.000 & 1.000 & 1.000&  1.000\\ 
\hline
\end{tabular}
\end{center}
\caption{\label{tab:nivel_mldsp5005} \footnotesize Frequency of rejection for the bootstrap test $T_{n,\ml}$ and  $T_{n,\ds}$ for $n=200$ and dimension $p=5$, $\alpha=0.05$.  {$\star$ indicates that the frequency of rejection is significantly different from the nominal level.}}
\end{table}
\normalsize

From Table \ref{tab:nivel_mldsp5005}, one observes that, except for the Pearson distribution,  the observed level of the classical procedure, $\pi_{H_0}(T_{n,\ml})$, is slightly higher than the nominal one.   {However,  $T_{n,\ml}$ leads to a liberal test only for the  ${\cal T}_{p,5}({\bf{0}}, \bI)$ distribution.} On the contrary,  $\pi_{H_0}(T_{n,\ds})$ is smaller than the nominal level,  except for the Cauchy distribution  in which case, the exact size is close to $0.05$.  {However, in none of the  considered situations the observed frequencies of rejection are significantly different from the nominal level $\alpha=0.05$}. For the Cauchy distribution, as expected, the classical test is non--informative when considering the alternatives $H_{1,\Delta}^{(4)}$. On the other hand, both procedures detect the alternatives $ H_1^{\star^{(j)}}$ for $j=1,\dots, 4$. Table \ref{tab:rho_mldsp5005} shows the advantage of the procedure based on $T_{n,\ds}$ over that based on $T_{n,\ml}$, except for $H_{1,0.05}^{(1)}$. The   inadequate behavior of $T_{n,\ml}$ for the Cauchy distribution shown by a power almost equal to the level for the alternatives  $H_{1,\Delta}^{(4)}$ is more clear when comparing the values of the size--corrected relative exact powers $\rho_{H_1}(T_{n,\ds},T_{n,\ml})$.

\footnotesize 
\begin{table}[ht!]
\begin{center}\footnotesize
\begin{tabular}{  c|  r r r r r r r r r}
  & \multicolumn{9}{c}{$\rho_{H_1}(T_{n,\ds},T_{n,\ml})$} \\\hline 
  &   \multicolumn{5}{c}{$H_{1,\Delta}^{(j)}$} & $H_1^{\star^{(1)}}$   & $H_1^{\star^{(2)}}$   & $H_1^{\star^{(3)}}$  & $H_1^{\star^{(4)}}$ \\\cline{2-6}
$\Delta$& $0.25$ &  $ 0.5$ & $0.75$ &   $ 1$ & $ 1.5$  & & & & \\\hline
$H_0^{(1)}$  & {-64.286} & -32.765 &{1.279} & 2.766 & 2.766 & 2.766 &2.766  &2.766  & 2.766     \\ 
$H_0^{(2)}$&   {-0.421} & 1.048  &{1.048} & 1.048   & 1.048 & 1.048  & 1.048  & 1.048  & 1.048   \\   
$H_0^{(3)}$ &  {-220.000} &  9.524 &{5.598} & 4.367  & 4.348 & 4.348 & 4.348 &4.348 & 4.348   \\ 
$H_0^{(4)}$ &  {-90.000} &  227.273 &{1128.571} & 1877.778  & 1466.667 &  0.425 & 0.425 &  0.425&   0.425\\ 
\hline
\end{tabular}
\end{center}
\caption{\label{tab:rho_mldsp5005} \footnotesize Size corrected relative exact power of the  robust bootstrap test $T_{n,\ds}$ with respect to the classical $T_{n,\ml}$ one, when  $n=200$, $p=5$ and $\alpha=0.05$.}
\end{table}
\normalsize

\subsection{Comparisons with other tests for elliptical symmetry}{\label{otros}}
Taking into account the better performance of the procedure based on  $T_{n,\ds}$ over that based on  $T_{n,\ml}$ and $T_{n,\ese}$, in this section, we compare the conditional test based on  $T_{n,\ds}$ with some other methods found in the literature. The simulation conditions are similar to those described in Sections \ref{montecarlo} and \ref{montecarlodim5}.

As mentioned in Section \ref{test}, there is a wide literature on methods to test for elliptical symmetry. According to the  simulation power studies performed in Huffer and Park (2007), none of the tests introduced in   Manzotti \textsl{et al.} (2002), Schott (2002) or Huffer and Park (2007) is uniformly superior for detecting departures from the null hypothesis. On the other hand, Batsidis \textsl{et al.} (2014) also showed that their proposal is comparable in power to that defined in  Schott (2002).

The purpose of the numerical study in this section, is to show that the proposed test statistic is a useful option to the previously defined methods, in particular when moments do not exist. Since there is no superior test statistic, we decided to choose for the comparison a test statistic which can be easily computed and has a tractable null distribution. For this reason, we excluded the test defined in  Koltchinskii  and Sakhanenko (2000) as well as the statistic defined in  Beran (1979). 
With respect to  the test proposed in Batsidis and Zografos (2013), their method   helps to decide departures from a specific elliptical model, while our procedure is designed   to detect departures from the whole family of elliptical distributions. For this reason, it is not included in the comparison.

On the contrary, the test defined by Schott (2002) is easy to compute since it is based on a fourth moment statistic denoted as  $T_{n,\sch}$. Besides, this statistic  is asymptotically $\chi^2_{\nu}$, where $\nu$ depends on the dimension of the data, but not on the underlying null elliptical distribution.  Based on the simulation studies reported in   Schott (2002), Huffer and Park (2007) and Batsidis \textsl{et al.} (2014), the test based on $T_{n,\sch}$ has observed level close to the nominal one and good empirical power. Besides, as our procedure, the test statistic is affine invariant. It is worth noting, that the asymptotic behavior of $T_{n,\sch}$ is derived for  distributions  having finite moments up to order eight so that it will be sensitive to departures from this assumption, even if the distribution is elliptical.

We also include in the comparison the test statistic, $T_{n,\bat}$, recently introduced   in Batsidis \textsl{et al.} (2014) that is based on a power divergence family of statistics depending on a parameter $\lambda$.   According to the simulation results in Batsidis \textsl{et al.} (2014), we select    $\lambda=1$ (similar results were obtained for  $\lambda=2/3$). As noted by these authors,  $T_{n,\bat}$ has approximately  a chi--square distribution with degrees of freedom depending on the sample size and may be computed in  a simple way.  As mentioned in Batsidis \textsl{et al.} (2014), this test statistic has a very good power for a variety of alternatives, even when it is not affine invariant.

The observed frequency of rejection and the size corrected relative exact powers for  $T_{n,\bat}$ are given in   Tables \ref{tab:nivelbat1} and    \ref{tab:rhobat1} for $p=2$ and in  Tables \ref{tab:nivelbat1p5} and   \ref{tab:rhobat1p5}, when $p=5$. Analogous quantities for $T_{n,\sch}$ are reported in Tables \ref{tab:nivelsch} and \ref{tab:rhosch} for $p=2$ and  in Tables \ref{tab:nivelschp5} and \ref{tab:rhoschp5} for $p=5$.

\footnotesize
\begin{table}[ht!]
\begin{center}\footnotesize
\begin{tabular}{  c| c c c c c c cc}
 & \multicolumn{8}{c}{$T_{n,\bat}$}\\\hline
    & & \multicolumn{3}{c}{$H_{1,\Delta}^{(j)}$}& $H_1^{\star^{(1)}}$ & $H_1^{\star^{(2)}}$ & $H_1^{\star^{(3)}}$ & $H_1^{\star^{(4)}}$\\\cline{3-5}
$\Delta$& $ 0$ & $ 0.5$ & $ 1$ & $ 1.5$  & & & &\\\hline
$H_0^{(1)}$  & 0.016{$\star$}& 0.048& 0.512& 0.980 & 0.456 & 1.000 & 0.974& 0.976\\ 
$H_0^{(2)}$&  0.570{$\star$} & 0.550 & 0.740 & 0.970 &  0.456 & 1.000 & 0.974& 0.976\\   
$H_0^{(3)}$ &   0.054 & 0.088 & 0.522 & 0.976 &  0.456 & 1.000  & 0.974& 0.976\\ 
$H_0^{(4)}$ &  0.140{$\star$} & 0.122 & 0.418 & 0.852   & 0.456 & 1.000  & 0.974& 0.976\\ 
$H_0^{(5)}$ &  0.014{$\star$} & 0.032 & 0.842 & 0.998   &  0.456 & 1.000  & 0.974& 0.976\\ 
$H_0^{(6)}$ &    0.030 & 0.832 & 1.000 & 1.000  & 0.456 & 1.000  & 0.974& 0.976\\ 
$H_0^{(7)}$ &   0.926{$\star$} & 0.892 & 0.904 & 0.892  &  0.456 & 1.000  & 0.974& 0.976\\ 
\hline
\end{tabular}
\end{center}
\caption{\label{tab:nivelbat1} \footnotesize Frequency of rejection for the  test defined in Batsidis \textsl{et al.} (2014) for $n=200$ and dimension $p=2$ with $\lambda=1$. {$\star$ indicates that the frequency of rejection is significantly different from the nominal level.}}
\end{table}
\normalsize

\footnotesize 
\begin{table}[ht!]
\begin{center}\footnotesize
\begin{tabular}{  c| r r r r r r r  }
&  \multicolumn{7}{c}{$\rho_{H_1}(T_{n,\bat},T_{n,\ds})$}\\\hline
    &   \multicolumn{3}{c}{$H_{1,\Delta}^{(j)}$}& $H_1^{\star^{(1)}}$ & $H_1^{\star^{(2)}}$ & $H_1^{\star^{(3)}}$ & $H_1^{\star^{(4)}}$\\\cline{2-4}
$\Delta$&   $ 0.5$ & $ 1$ & $ 1.5$  & & & & \\\hline
$H_0^{(1)}$  &  -84.615  & -47.346  &  1.261 & -53.782 &  3.361  &  0.630   & 0.840  \\ 
$H_0^{(2)}$&   -108.889 & -81.642 & -57.537  &-112.102  &-54.352   & -57.113& -56.900 \\   
$H_0^{(3)}$ &   -83.168  &-50.319 & -3.758 & -58.038 & -1.253  & -3.967& -3.758\\ 
$H_0^{(4)}$ &  -112.000  & -65.509&  -24.416  & -66.525  & -8.898   & -11.653& -11.441\\ 
$H_0^{(5)}$ &  -96.703 & -11.727   & 4.904  & -52.879 &  5.117    &  2.345 & 2.559\\ 
$H_0^{(6)}$ &   -15.756  &  1.891 &  1.891 & -55.252   & 1.891  & -0.840& -0.630\\ 
$H_0^{(7)}$ &  -288.889 &-105.289& -104.337 &-149.370 & -92.227 & -94.958  & -94.748\\ 
\hline
\end{tabular}
\end{center}
\caption{\label{tab:rhobat1} \footnotesize Size corrected relative exact power for test based on the statistic $T_{n,\bat}$ defined in  Batsidis \textsl{et al.} (2014) with respect to the robust bootstrap test  $T_{n,\ds}$   for $n=200$ and dimension $p=2$ with $\lambda=1$.}
\end{table}
\normalsize

\footnotesize 
\begin{table}[ht!]
\begin{center}\footnotesize
\begin{tabular}{  c| c c c c c c c c cc}
  & \multicolumn{10}{c}{$T_{n,\bat}$}\\\hline
    & & \multicolumn{5}{c}{$H_{1,\Delta}^{(j)}$} & $H_1^{\star^{(1)}}$  & $H_1^{\star^{(2)}}$   & $H_1^{\star^{(3)}}$   & $H_1^{\star^{(4)}}$\\\cline{3-7}
$\Delta$& $ 0$ & $0.25$ & $ 0.5$ & $0.75$ & $ 1$ & $ 1.5$  & & \\\hline
$H_0^{(1)}$  &  0.026  & {0.030} & 0.060 &{0.186} & 0.630  & 0.998 & 0.936   &1.000  &0.888& 0.990\\  
$H_0^{(2)}$&   0.020{$\star$}  & {0.080} & 0.970  &{1.000} &1.000  &1.000  & 0.936  & 1.000 & 0.888& 0.990\\    
$H_0^{(3)}$ &  0.036 & {0.036} & 0.050 &{0.210} & 0.516  &0.972  & 0.936    & 1.000 & 0.888& 0.990 \\   
$H_0^{(4)}$ &  0.800{$\star$}  & {0.804} & 0.798 &{0.808} &0.810  &0.846  & 0.936    &1.000  &0.888& 0.990\\  
\hline
\end{tabular}
\end{center}
\caption{\label{tab:nivelbat1p5} \footnotesize Frequency of rejection for the  test defined in Batsidis \textsl{et al.} (2014) for $n=200$ and dimension $p=5$ with $\lambda=1$. {$\star$ indicates that the frequency of rejection is significantly different from the nominal level.}}
\end{table}
\normalsize

\footnotesize 
\begin{table}[ht!]
\begin{center}\footnotesize
\begin{tabular}{  c|  r r r r r r r r r }
& \multicolumn{9}{c}{$\rho_{H_1}(T_{n,\bat},T_{n,\ds})$}\\\hline
   &   \multicolumn{5}{c}{$H_{1,\Delta}^{(j)}$} & $H_1^{\star^{(1)}}$   & $H_1^{\star^{(2)}}$   & $H_1^{\star^{(3)}}$  & $H_1^{\star^{(4)}}$ \\\cline{2-6}
$\Delta$&  $0.25$ & $ 0.5$ & $0.75$ & $ 1$ & $ 1.5$  & & & & \\\hline
$H_0^{(1)}$  &  {-60.000} & -91.371 &{-83.158} & -37.474   & 0.621  &  -5.797    & 0.828  & -10.766    & -0.207   \\ 
$H_0^{(2)}$&   {-93.658} &-1.452 & {1.660} & 1.660 & 1.660 & -4.979   & 1.660 &-9.959 & 0.622 \\   
$H_0^{(3)}$ &   {-100.000} & -93.913 & {-79.036}  & -49.791 & -2.500   &-6.250   & 0.417& -11.250   & -0.625  \\ 
$H_0^{(4)}$ &  {100.000} & -102.778 & {-97.674} &-98.596 & -95.106  & -85.624   & -78.858  & -90.698&-79.915 
\\ 
\hline
\end{tabular}
\end{center}
\caption{\label{tab:rhobat1p5} \footnotesize Size corrected relative exact power for test based on the statistic $T_{n,\bat}$ defined in  Batsidis \textsl{et al.} (2014) with respect to the robust bootstrap test  $T_{n,\ds}$   for $n=200$ and dimension $p=5$ with $\lambda=1$.}
\end{table}
 \normalsize

\footnotesize
\begin{table}[ht!]
\begin{center}\footnotesize
\begin{tabular}{  c| c c c c c c cc}
  & \multicolumn{8}{c}{$T_{n,\sch}$}\\\hline
     & & \multicolumn{3}{c}{$H_{1,\Delta}^{(j)}$}& $H_1^{\star^{(1)}}$ & $H_1^{\star^{(2)}}$ & $H_1^{\star^{(3)}}$ & $H_1^{\star^{(4)}}$\\\cline{3-7}
$\Delta$& $ 0$ & $ 0.5$ & $ 1$ & $ 1.5$  & & \\\hline
$H_0^{(1)}$  & 0.032  & 0.074  & 0.218  & 0.312  & 0.548    & 0.282  & 0.140 & 0.864\\ 
$H_0^{(2)}$&  0.504{$\star$}  & 0.490  & 0.444  & 0.394   & 0.548    & 0.282  & 0.140& 0.864\\   
$H_0^{(3)}$ &  0.054  & 0.050  & 0.182  & 0.296  & 0.548    & 0.282  & 0.140& 0.864\\ 
$H_0^{(4)}$ & 0.098{$\star$}  & 0.094  & 0.094  & 0.158   & 0.548   & 0.282  & 0.140& 0.864\\ 
$H_0^{(5)}$ & 0.064  & 0.112  & 0.310  & 0.380   & 0.548    & 0.282  & 0.140& 0.864\\ 
$H_0^{(6)}$ &   0.054  & 0.276  & 0.382  & 0.410   & 0.548     & 0.282  & 0.140& 0.864 \\ 
$H_0^{(7)}$ & 0.396{$\star$}  & 0.394  & 0.410  & 0.432   & 0.548   & 0.282  & 0.140& 0.864\\ 
\hline
\end{tabular}
\end{center}
\caption{\label{tab:nivelsch} \footnotesize Frequency of rejection for the  test defined in Schott (2002) for $n=200$ and dimension $p=2$. {$\star$ indicates that the frequency of rejection is significantly different from the nominal level.}}
\end{table}
 \normalsize

\footnotesize
\begin{table}[ht!]
\begin{center}\footnotesize
\begin{tabular}{  c|  r r r r r r r}
&   \multicolumn{7}{c}{$\rho_{H_1}(T_{n,\sch},T_{n,\ds})$}\\\hline
    &   \multicolumn{3}{c}{$H_{1,\Delta}^{(j)}$}& $H_1^{\star^{(1)}}$ & $H_1^{\star^{(2)}}$ & $H_1^{\star^{(3)}}$ & $H_1^{\star^{(4)}}$\\\cline{2-4}
$\Delta$&  $ 0.5$ & $ 1$ & $ 1.5$  & & & &\\\hline
$H_0^{(1)}$  &  -79.808 & -80.255 & -70.588  & -45.798   & -73.740 & -88.656& -12.605 \\ 
$H_0^{(2)}$&   -106.222 & -106.480 & -111.677   & -95.329   & -123.567 & -138.641  & -61.783 \\   
$H_0^{(3)}$ &   -101.980   & -86.412  & -74.739   & -48.434   & -76.200   & -91.023  & -15.449\\ 
$H_0^{(4)}$ &   -102.667 & -100.496   & -93.631  & -52.331    & -80.509  & -95.551& -18.856\\ 
$H_0^{(5)}$ &   -91.209 & -73.774  & -66.311  & -48.401 & -76.759  & -91.898 & -14.712 \\ 
$H_0^{(6)}$ &   -76.681 & -65.546  & -62.605  & -48.109   & -76.050  & -90.966 & -14.916 \\ 
$H_0^{(7)}$ &   -111.111   & -96.635  & -95.408   & -84.034  & -111.975 & -126.891 & -50.840 \\ 
\hline
\end{tabular}
\end{center}
\caption{\label{tab:rhosch} \footnotesize Size corrected relative exact power for test based on the statistic $T_{n,\sch}$ defined in  Schott (2002) with respect to the robust bootstrap test  $T_{n,\ds}$   for $n=200$ and dimension $p=2$.}
\end{table}
 \normalsize

\footnotesize
\begin{table}[ht!]
\begin{center}\footnotesize
\begin{tabular}{  c| c c c c c c c c cc}
 & \multicolumn{10}{c}{$T_{n,\sch}$}\\\hline
   & & \multicolumn{5}{c}{$H_{1,\Delta}^{(j)}$} & $H_1^{\star^{(1)}}$  & $H_1^{\star^{(2)}}$   & $H_1^{\star^{(3)}}$   & $H_1^{\star^{(4)}}$\\\cline{3-5}
$\Delta$& $ 0$ & $0.25$ &  $ 0.5$ & $0.75$ &$ 1$ & $ 1.5$  & & \\\hline
$H_0^{(1)}$  & 0.040  & {0.064} & 0.182  & {0.550} & 0.790  & 0.900  &  0.864 & 0.872 & 0.314 & 0.516 \\  
$H_0^{(2)}$&   0.056  & {0.468} & 0.900  &  {0.944} & 0.954  & 0.964  & 0.864 & 0.872 & 0.314 & 0.516\\    
$H_0^{(3)}$ &  0.054  & {0.046} & 0.052  & {0.150} & 0.410  & 0.800  & 0.864 & 0.872 & 0.314 & 0.516\\   
$H_0^{(4)}$ &  0.974{$\star$}  & {0.974} & 0.974 &{0.974} & 0.970  & 0.946  & 0.864 & 0.872 & 0.314 & 0.516\\  
\hline
\end{tabular}
\end{center}
\caption{\label{tab:nivelschp5} \footnotesize Frequency of rejection for the  test defined in Schott (2002) for $n=200$ and dimension $p=5$. {$\star$ indicates that the frequency of rejection is significantly different from the nominal level.}}
\end{table}
\normalsize

\footnotesize
\begin{table}[ht!]
\begin{center}\footnotesize
\begin{tabular}{  c|  r r r r r r r r r}
  & \multicolumn{9}{c}{$\rho_{H_1}(T_{n,\sch},T_{n,\ds})$} \\\hline 
  &   \multicolumn{5}{c}{$H_{1,\Delta}^{(j)}$} & $H_1^{\star^{(1)}}$   & $H_1^{\star^{(2)}}$   & $H_1^{\star^{(3)}}$  & $H_1^{\star^{(4)}}$ \\\cline{2-6}
$\Delta$&  $0.25$ & $ 0.5$ &$0.75$ & $ 1$ & $ 1.5$  & & & & \\\hline
$H_0^{(1)}$  &{140.000} &-63.959 & {-46.316}   & -22.360  & -10.973  & -14.700  & -13.872  &-71.636 &  -50.725  \\ 
$H_0^{(2)}$&   {-56.448} & -12.448 & {-7.884}    & -6.847  & -5.809   &-16.183   &-15.353  &-73.237 &-52.282  \\   
$H_0^{(3)}$ &  {-166.667} &   -100.870 & {-88.434}    &-62.762  & -22.292  & -15.625     &-14.792   &-72.917  &-51.875  \\ 
$H_0^{(4)}$ & {-100.000} & -100.000  & {-100.000}   &-100.562  &-102.979  &-111.628  &-110.782  & -169.767 & -148.414  \\ 
\hline
\end{tabular}
\end{center}
\caption{\label{tab:rhoschp5} \footnotesize Size corrected relative exact power for test based on the statistic $T_{n,\sch}$ defined in Schott (2002) with respect to the robust bootstrap test  $T_{n,\ds}$   for $n=200$ and dimension $p=5$.}
\end{table}

\normalsize

 As expected, when the underlying distribution has no moments, the test based on $T_{n,\sch}$ becomes non informative, since it relies on the assumption of existence of eight order moments. The same happens when the data are generated according to the multivariate Student distributions ${\cal T}_{2,3}(\mbox{\bf 0}, \identidad)$ and ${\cal T}_{5,5}(\mbox{\bf 0}, \identidad)$ since these distributions do not satisfy the assumptions in  Schott (2002). When $p=2$, in the four other situations,  even if the test reaches in most cases the desired level, it has difficulties to detect the selected alternatives, specially the alternatives $H_1^{\star^{(3)}}$ and $H_1^{\star^{(4)}}$. In dimension $p=5$,  for the Pearson distribution, the level and power performance of $T_{n,\sch}$ is analogous to that of our procedure, while for the normal distribution, the test based on $T_{n,\ds}$ has a better detection power except for $\Delta=0.25$. For the chosen fixed alternatives, $H_1^{\star^{(j)}}$, $j=3, 4$ our procedure has much better power.

 With respect to the proposal given in Batsidis \textsl{et al.} (2014), the test based on $T_{n,\bat}$ becomes non--informative when contaminating with a Cauchy distribution or when the data follow a Cauchy distribution, which can be explained by the fact that the test is based on the sample mean and the sample covariance matrix. On the other hand, for the number of replications considered the level $0.05$ is not attained for the normal distribution and for data uniformly distributed over the unit circle, in dimension $p=2$. This fact was also observed in Table 10 of Batsidis \textsl{et al.} (2014) for the normal distribution with 10000 replications. Table \ref{tab:rhobat1}  shows that, as for the Schott's test,  our method outperforms the procedure based on $T_{n,\bat}$  {in the majority of the cases considered}. Similar conclusions   {for the considered distributions and alternatives} are obtained in dimension $p=5$, as shown in Table    \ref{tab:nivelbat1p5}.
 
 These facts are highlighted in Tables  \ref{tab:rhosch} and \ref{tab:rhoschp5} that report the size corrected relative powers $\rho_{H_1}(T_{n,\sch},T_{n,\ds})$. The negative values reported in all cells confirm the better performance of $T_{n,\ds}$. Note that  {even if}, for the Cauchy distribution, the test proposed in  Batsidis \textsl{et al.} (2014) has a positive value when $\Delta=0.25$ and $p=5$, the test is non--informative having power almost constant for $H_{1,\Delta}^{(4)}$.  {Hence, in this case, the size corrected size  does not provide a good measure to compare the test statistics}.
 
  {It is worth noticing that our conclusions regarding the better performance of the test based on the Donoho--Stahel estimators are valid only for the considered distributions and alternatives.  A more extensive simulation study would be necessary to conclude that, in general,  $T_{n,\ds}$ should be preferred. This interesting comparison may be object of future work.}

 \subsection{Simulation study in dimension $p=5$ with $n=50$.}{\label{dim5n50}}
In Section \ref{numerical}, we noticed that for small sample sizes the distribution of the test statistic does not allow to distinguish the elliptical distributions from those considered in the alternative. For that reason, in the simulation study reported in Sections \ref{montecarlo} and \ref{montecarlodim5}, we choose as sample size $n=200$. To complement the results obtained in Section \ref{montecarlodim5} and to study the effect of a smaller sample size on the decisions taken, we report here the observed frequencies of rejection for the   test based on the sample mean and covariance matrix, $T_{n,\ml}$ and for that based on the Donoho--Stahel estimators, $T_{n,\ds}$, when $p=5$ and $n=50$, which represents a challenging situation due to the ratio between sample size and dimension. As in Section \ref{otros}, we also compare their performance with that of the   test statistic, $T_{n,\bat}$,   introduced   in Batsidis \textsl{et al.} (2014)  and with the test defined by Schott (2002), $T_{n,\sch}$. The simulation conditions as well as the considered alternatives are described in Section \ref{montecarlodim5}.
 
{The corresponding frequencies of rejection are reported in Tables \ref{tab:nivel_mldsp5005_n50}, \ref{tab:nivelbat1p5_n50} and \ref{tab:nivelschp5_n50}, where $\Delta=0$ corresponds to the observations generated according to the null hypothesis. Besides, Table \ref{tab:rho_mldsp5005_n50} reports the size--corrected relative exact powers $\rho_{H_1}(T_{n,\ds},T_{n,\ml})$ as defined in (7), while Tables \ref{tab:rhobat1p5_n50} and \ref{tab:rhoschp5_n50}  report the size corrected relative powers $\rho_{H_1}(T_{n,\bat},T_{n,\ds})$ and $\rho_{H_1}(T_{n,\sch},T_{n,\ds})$. Note that a positive value of size corrected relative power $\rho_{H_1}(T_{n,1},T_{n,2})$ indicates that the test based on $T_{n,1}$ has a better detection capability than that based on $T_{n,2}$ and the size of its advantage is quantified by  $\rho_{H_1}(T_{n,1},T_{n,2})$. Similarly, a negative value of  $\rho_{H_1}(T_{n,1},T_{n,2})$ provides a measure of the deficiency of $T_{n,1}$ with respect to $T_{n,2}$.}

\footnotesize 
 
\begin{table}[ht!]
\begin{center}\footnotesize
\begin{tabular}{  c| c cc c c c c c c c}
  & \multicolumn{10}{c}{$T_{n,\ml}$} \\\hline 
  & & \multicolumn{5}{c}{$H_{1,\Delta}^{(j)}$} & $H_1^{\star^{(1)}}$  & $H_1^{\star^{(2)}}$   & $H_1^{\star^{(3)}}$   & $H_1^{\star^{(4)}}$ \\\cline{3-7}
$\Delta$& $ 0$ & $0.25$ & $ 0.5$ & $0.75$ & $ 1$ & $ 1.5$  & & & &\\\hline
$H_0^{(1)}$  &  0.054 & 0.068 & 0.180 & 0.528 & 0.844 & 0.988  & 1.000  & 1.000  & 1.000  & 0.998\\ 
 
$H_0^{(2)}$& 0.044 & 0.440 & 0.984 & 1.000 & 1.000 &   1.000 & 1.000  & 1.000 &  1.000  & 0.998 \\   
$H_0^{(3)}$ & 0.128$\star$  & 0.134 & 0.186 & 0.370 & 0.652 & 0.940   & 1.000  & 1.000 &  1.000  & 0.998\\ 
$H_0^{(4)}$ & 0.228$\star$ & 0.230 & 0.240 & 0.276 & 0.290 & 0.362 &  1.000  & 1.000 &  1.000 &  0.998 \\ 
\hline
  & \multicolumn{10}{c}{$T_{n,\ds}$} \\\hline 
  & & \multicolumn{5}{c}{$H_{1,\Delta}^{(j)}$} & $H_1^{\star^{(1)}}$  & $H_1^{\star^{(2)}}$   & $H_1^{\star^{(3)}}$  & $H_1^{\star^{(4)}}$  \\\cline{3-7}
$\Delta$& $ 0$ & $0.25$ & $ 0.5$ & $0.75$ & $ 1$ & $ 1.5$  & & & &\\\hline
 $H_0^{(1)}$  & 0.052 & 0.050 & 0.086&   0.236 & 0.542 & 0.898   &  0.992  &  1.000 &   0.972  &  0.980\\ 
 $H_0^{(2)}$&  0.060 & 0.204 & 0.858 & 0.992 & 1.000 & 1.000 &  0.992  &  1.000 &   0.972  &  0.980\\   
$H_0^{(3)}$ & 0.068 & 0.050 & 0.090 & 0.190 & 0.360 & 0.774  &  0.992  &  1.000 &   0.972  &  0.980\\ 
$H_0^{(4)}$ &  0.044 & 0.036 & 0.048 & 0.098 & 0.150 & 0.330  &  0.992  &  1.000 &   0.972  &  0.980\\ 
\hline
\end{tabular}
\end{center}
\caption{\label{tab:nivel_mldsp5005_n50} \footnotesize Frequency of rejection for the bootstrap test $T_{n,\ml}$ and  $T_{n,\ds}$ for $n=50$ and dimension $p=5$, $\alpha=0.05$. $\star$ indicates that the frequency of rejection is significantly different from the nominal level.}
\end{table}
 
\normalsize
\footnotesize 
 
\begin{table}[ht!]
\begin{center}\footnotesize
\begin{tabular}{  c| c c c c c c c c cc}
  & \multicolumn{10}{c}{$T_{n,\bat}$}\\\hline
    & & \multicolumn{5}{c}{$H_{1,\Delta}^{(j)}$} & $H_1^{\star^{(1)}}$  & $H_1^{\star^{(2)}}$   & $H_1^{\star^{(3)}}$   & $H_1^{\star^{(4)}}$\\\cline{3-7}
$\Delta$& $ 0$ & $0.25$ & $ 0.5$ & $0.75$ & $ 1$ & $ 1.5$  & & \\\hline
$H_0^{(1)}$  &0.016$\star$  & 0.016 & 0.030  & 0.040 & 0.084 & 0.248 & 0.152 & 0.302 & 0.144 & 0.274\\  
$H_0^{(2)}$&  0.016$\star$  & 0.036 & 0.186 & 0.440  & 0.544 & 0.640  & 0.152 & 0.302 & 0.144 & 0.274\\    
$H_0^{(3)}$ & 0.016$\star$ &  0.024 &  0.036 &  0.036 &  0.074 &  0.184 & 0.152 & 0.302 & 0.144 & 0.274 \\   
$H_0^{(4)}$ &   0.438$\star$ & 0.408&  0.362&  0.374 & 0.352 &  0.324 & 0.152 & 0.302 & 0.144 & 0.274\\  
\hline
\end{tabular}
\end{center}
\caption{\label{tab:nivelbat1p5_n50} \footnotesize Frequency of rejection for the  test defined in Batsidis \textsl{et al.} (2014) for $n=50$ and dimension $p=5$ with $\lambda=1$.  $\star$ indicates that the frequency of rejection is significantly different from the nominal level.}
\end{table}
 
\normalsize

\footnotesize
 
\begin{table}[ht!]
\begin{center}\footnotesize
\begin{tabular}{  c| c c c c c c c c cc}
 & \multicolumn{10}{c}{$T_{n,\sch}$}\\\hline
   & & \multicolumn{5}{c}{$H_{1,\Delta}^{(j)}$} & $H_1^{\star^{(1)}}$  & $H_1^{\star^{(2)}}$   & $H_1^{\star^{(3)}}$   & $H_1^{\star^{(4)}}$\\\cline{3-5}
$\Delta$& $ 0$ & $0.25$ &  $ 0.5$ & $0.75$ &$ 1$ & $ 1.5$  & & \\\hline
$H_0^{(1)}$  &  0.048 & 0.034 & 0.034 & 0.076 &  0.142 & 0.312& 0.238 & 0.258 & 0.074 & 0.708  \\  
$H_0^{(2)}$&   0.060 & 0.074 & 0.270 & 0.432&  0.514 & 0.586 & 0.238 & 0.258 & 0.074 & 0.708 \\    
$H_0^{(3)}$ & 0.038 & 0.036 & 0.044 & 0.060 & 0.084 & 0.182  & 0.238 & 0.258 & 0.074 & 0.708 \\   
$H_0^{(4)}$ &  0.964$\star$ & 0.966&  0.966 & 0.956 & 0.948 & 0.900 & 0.238 & 0.258 & 0.074 & 0.708 \\  
\hline
\end{tabular}
\end{center}
\caption{\label{tab:nivelschp5_n50} \footnotesize Frequency of rejection for the  test defined in Schott (2002) for $n=50$ and dimension $p=5$. $\star$ indicates that the frequency of rejection is significantly different from the nominal level.}
\end{table}
 
\normalsize

{Table  \ref{tab:nivel_mldsp5005_n50} shows that, even for this small sample size, the robust procedures  allow to detect the considered alternatives keeping the  exact sizes of the test statistic, i.e.,  $\pi_{H_0}(T_{n,\ds})$ around the nominal level $\alpha=0.05$. As in Section \ref{montecarlodim5} and \ref{otros}, we indicate with a $\star$ those cases, in which the observed empirical frequencies of rejection are   different from the nominal level with a significance level $\gamma=0.01$. As expected, the test based on the sample mean and covariance matrix outperforms that based on the Donoho--Stahel estimators under the normal distribution due to the loss of efficiency of the robust estimators. The advantage of $T_{n,\ml}$ is also observed for the Pearson distribution, in particular, when $\Delta=0.25$ and $\Delta=0.5$. These two facts are consistent with the behaviour described in Section \ref{numerical}, where for dimension $p=2$ and $n=50$, the distribution of the test statistic has troubles to distinguish between the null hypothesis and close alternatives for most of the considered elliptical distributions. Note that the test based on the sample mean and covariance matrix becomes liberal under ${\cal T}_{p,5}({\bf{0}}, \bI)$. Besides,  the test based on the robust estimators shows its advantage for this distribution, except for $\Delta=0.25$ where $T_{n,\ds}$ does not succeed in detecting the hypothesis, leading to a large negative value on the size corrected relative power. On the other hand, under the Cauchy distribution $T_{n,\ml}$ is non--informative, while  $T_{n,\ds}$ is able to distinguish all the alternatives except when $\Delta= 0.25$ and $0.5$. These facts become more evident in Table \ref{tab:rho_mldsp5005_n50}, where most size corrected relative powers are positive for distributions different from the normal. The large negative value obtained at $\Delta=0.25$ for the Cauchy distribution can be explained by means of two facts. The first one is that  $T_{n,\ml}$ has power almost constant, so that the denominator is close to 0, while the second one is that the power of  $T_{n,\ds}$ decreases   at $\Delta=0.25$ with respect to its size. Note that, given two test statistics $T_{n,1}$ and $T_{n,2}$, when the test based on $T_{n,2}$ is non-informative, a negative value of the size corrected relative power $\rho_{H_1}(T_{n,1},T_{n,2})$  does not provide a good measure to conclude the benefits of $T_{n,2}$ over $T_{n,1}$.  }

{With respect to the test statistics, $T_{n,\bat}$   and  $T_{n,\sch}$, both procedures loose their capability of detection under the Cauchy distribution, since their behaviour relies on the existence of moments. On the other hand, the test statistic proposed    in Batsidis \textsl{et al.} (2014) is conservative in all situations except for the Cauchy distribution ($H_0^{(4)}$), where it is liberal. The same conclusions obtained when $n=200$ are preserved in the actual setting, that is, the procedure proposed in this paper outperforms these competitors in the majority of the situations considered. As mentioned in Section \ref{otros},   our conclusions on the benefits of  $T_{n,\ds}$ are valid only for the considered distributions and alternatives. Quite surprisingly, even for this small sample size the procedure based on $T_{n,\ds}$ shows a reasonable performance probably due to the bootstrap method used to compute the $p-$value.}

\footnotesize 
 
\begin{table}[ht!]
\begin{center}\footnotesize
\begin{tabular}{  c|  r r r r r r r r r}
  & \multicolumn{9}{c}{$\rho_{H_1}(T_{n,\ds},T_{n,\ml})$} \\\hline 
  &   \multicolumn{5}{c}{$H_{1,\Delta}^{(j)}$} & $H_1^{\star^{(1)}}$   & $H_1^{\star^{(2)}}$   & $H_1^{\star^{(3)}}$  & $H_1^{\star^{(4)}}$ \\\cline{2-6}
$\Delta$& $0.25$ &  $ 0.5$ & $0.75$ &   $ 1$ & $ 1.5$  & & & & \\\hline
$H_0^{(1)}$  & -114.286    & -73.016  & -61.181 &  -37.975 &  -9.422  & -0.634 &   0.211 &  -2.748 & -1.695 \\ 
$H_0^{(2)}$&    -63.636 & -15.106  & -2.510 & -1.674  & -1.674 & -2.510 & -1.674 & -4.603 & -3.564   \\   
$H_0^{(3)}$ &    -400.000 & -62.069 & -49.587 & -44.275 & -13.054  &  5.963  &  6.881  &  3.670   & 4.828  \\ 
$H_0^{(4)}$ &  -500.000 &  -66.667 &  12.500 &  70.968  & 113.433  & 22.798 &  23.834  & 20.207  & 21.548   \\ 
\hline
\end{tabular}
\end{center}
\caption{\label{tab:rho_mldsp5005_n50} \footnotesize Size corrected relative exact power of the  robust bootstrap test $T_{n,\ds}$ with respect to the classical $T_{n,\ml}$ one, when  $n=50$, $p=5$ and $\alpha=0.05$.}
\end{table}
 
\normalsize

\footnotesize 
 
\begin{table}[ht!]
\begin{center}\footnotesize
\begin{tabular}{  c|  r r r r r r r r r }
& \multicolumn{9}{c}{$\rho_{H_1}(T_{n,\bat},T_{n,\ds})$}\\\hline
   &   \multicolumn{5}{c}{$H_{1,\Delta}^{(j)}$} & $H_1^{\star^{(1)}}$   & $H_1^{\star^{(2)}}$   & $H_1^{\star^{(3)}}$  & $H_1^{\star^{(4)}}$ \\\cline{2-6}
$\Delta$&  $0.25$ & $ 0.5$ & $0.75$ & $ 1$ & $ 1.5$  & & & & \\\hline
$H_0^{(1)}$  &  -100.000 & -58.824   & -86.957 &  -86.122 & -72.577 & -85.532 & -69.831 & -86.087 & -72.198 \\ 
$H_0^{(2)}$&   -86.111 & -78.697 & -54.506 & -43.830 & -33.617 & -85.408 & -69.574 & -85.965 & -71.957  \\   
$H_0^{(3)}$ &    -144.444  & -9.091 &  -83.607 & -80.137 & -76.204 & -85.281 & -69.313 & -85.841 & -71.711 \\ 
$H_0^{(4)}$ &   275.000 & -2000.000  & -218.519  & -181.132 & -139.860 &  -130.169 & -114.226 & -131.681 & -117.949 \\ 
\hline
\end{tabular}
\end{center}
\caption{\label{tab:rhobat1p5_n50} \footnotesize Size corrected relative exact power for test based on the statistic $T_{n,\bat}$ defined in  Batsidis \textsl{et al.} (2014) with respect to the robust bootstrap test  $T_{n,\ds}$   for $n=50$ and dimension $p=5$ with $\lambda=1$.}
\end{table}
 
 \normalsize

\footnotesize
 
\begin{table}[ht!]
\begin{center}\footnotesize
\begin{tabular}{  c|  r r r r r r r r r}
  & \multicolumn{9}{c}{$\rho_{H_1}(T_{n,\sch},T_{n,\ds})$} \\\hline 
  &   \multicolumn{5}{c}{$H_{1,\Delta}^{(j)}$} & $H_1^{\star^{(1)}}$   & $H_1^{\star^{(2)}}$   & $H_1^{\star^{(3)}}$  & $H_1^{\star^{(4)}}$ \\\cline{2-6}
$\Delta$&  $0.25$ & $ 0.5$ &$0.75$ & $ 1$ & $ 1.5$  & & & & \\\hline
$H_0^{(1)}$  & 600.000  & -141.176   & -84.783  &  -80.816 &   -68.794 & -79.787 & -77.848 & -97.174 & -28.879 \\ 
$H_0^{(2)}$&   -90.278 & -73.684 & -60.086 & -51.702 & -44.043 & -80.901 & -78.936 & -98.465 & -29.565  \\   
$H_0^{(3)}$ &   -88.889 & -72.727 & -81.967 & -84.247 & -79.603 & -78.355 & -76.395 & -96.018 & -26.535 \\ 
$H_0^{(4)}$ &  -125.000 & -50.000 & -114.815 & -115.094 & -122.378 & -176.582 & -173.849 & -195.905 & -127.350 \\ 
\hline
\end{tabular}
\end{center}
\caption{\label{tab:rhoschp5_n50} \footnotesize Size corrected relative exact power for test based on the statistic $T_{n,\sch}$ defined in Schott (2002) with respect to the robust bootstrap test  $T_{n,\ds}$   for $n=50$ and dimension $p=5$.}
\end{table}
 
\normalsize

\small
 \noi \textbf{\small Acknowledgement} {\small  This research was   partially supported by Grants W276 and  20120130100241\textsc{ba} from Universidad of Buenos Aires, \textsc{pip}   112-2011-01-00339 from \textsc{conicet} and \textsc{pict} 2011-0397 from \textsc{anpcyt}, Argentina and also received financial support from Portuguese National Funds through FCT (Funda\c{c}\~{a}o para a Ci\^{e}ncia e a Tecnologia) under the scope of project PEst-OE/MAT/UI0822/2011. The authors wish to thank  two anonymous referees for valuable comments which led to an improved version of the original paper.}

\normalsize

{\setcounter{equation}{0}
\renewcommand{\theequation}{A.\arabic{equation}}
{\setcounter{section}{0}
\renewcommand{\thesection}{\Alph{section}}

\section*{Appendix A: Numerical computation of the test statistic}{\label{app}}
The defined  test statistic $T_{n, \bm,\bV }$  involves an integral that may be calculated numerically.
In dimension 2, the approximation described below is easy to perform. Assume that $w$ has compact support  $\itI=[-b,b]$   and   split it in a grid of ${N}_\itI$ points $t_i$.  
We consider ${ M}$ random directions $\{\ba_j\}_{j=1}^{ M}$ in  $\itS_p$ generated according to a uniform distribution on the sphere. 
Once the robust estimates $\bm_n$ and $\bV_n$  are obtained from the sample, for each $t_i$ on the grid and each generated random direction $\ba_j$, we compute  $I_{ij}=\left \lbrace \sqrt{n}P_n  \sin\left[ t_i \ba_j\trasp \bV_n^{-1/2} \left( \bX-\bm_n \right) \right]  \right \rbrace^2w(t_i)$. Then, we approximate the desired test statistic by $ {2 b \,   sur(\itS_p) }\, \sum_{i=1}^{{N}_\itI} \sum_{j=1}^{M} I_{ij} /({{N}_\itI {M} })$,
 where $sur(\itS_p)$ denotes the surface area of the sphere in $\real^p$ of radius $1$. 

To get an alternative expression for the test statistic,  we will restrict our attention to the situation where $w(t)=\indica_{[-b,b]}(t)/(2b)$. Let $Z_i(\ba)= \ba\trasp \bV_n^{-1/2} \left( \bX_i-\bm_n \right)$, then
$$
 \int    \left(\sqrt{n}P_n\left \lbrace \sin\left( t Z_i(\ba)\right) \right\rbrace  \right)^2\;w(t)dt
  =\frac{1}{2\,b\,n} \sum_{i,j}\int_{-b}^b   \sin\left( t Z_i(\ba) \right) \sin\left( t Z_j(\ba) \right) dt\;.
$$
Using that $\sin(x)\sin(y)=(\cos(x-y)-\cos(x+y))/2$ and denoting  $U_{ij}^{+}(\ba)=Z_i(\ba)+Z_j(\ba)=\ba\trasp\bV_n^{-1/2} \left(\bX_i+\bX_j-2\bm_n\right)$ and $U_{ij}^{-}(\ba)=Z_i(\ba)-Z_j(\ba)=\ba\trasp\bV_n^{-1/2} \left(\bX_i-\bX_j\right)$, we get that
\begin{eqnarray*}
 \int    \left(\sqrt{n}P_n\left \lbrace \sin\left( t Z_i(\ba)\right) \right\rbrace  \right)^2\;w(t)dt
 &=&\frac{1}{2\,b} \frac{1}{2\,n}\sum_{i,j}\int_{-b}^b   \cos\left( t U_{ij}^{-}(\ba) \right)  dt -\int_{-b}^b   \cos\left( t U_{ij}^{+}(\ba) \right)  dt\\
 &=&\frac{1}{2\,b\,n} \sum_{i,j} \frac{\sin\left( b U_{ij}^{-}(\ba) \right)}{U_{ij}^{-}(\ba) }  -  \frac{\sin\left( b U_{ij}^{+}(\ba) \right)}{U_{ij}^{+}(\ba)}\;,
  \end{eqnarray*}
which implies that
$$
T_{n, \bm,\bV }=\frac{1}{2\,n} \  \sum_{i,j}  \left[\esp_{\itU_p} \frac{\sin\left( b\, \bU\trasp\bV_n^{-1/2} \left(\bX_i-\bX_j\right) \right)}{b\,\bU\trasp\bV_n^{-1/2} \left(\bX_i-\bX_j\right) }  -  \esp_{\itU_p} \frac{\sin\left( b \,\bU\trasp\bV_n^{-1/2} \left(\bX_i+\bX_j-2\bm_n\right) \right)}{b\,\bU\trasp\bV_n^{-1/2} \left(\bX_i+\bX_j-2\bm_n\right)}\right] \,,
$$
where $\bU=(U_1,\dots,U_p)\trasp \sim \itU(\itS_p)=\itU_p$. 
Define $(D_{ij}^{+})^2=d^2(\bX_i-\bmu,-(\bX_j-\bmu), \bV)$ and $
(D_{ij}^{-})^2=d^2(\bX_i-\bmu,\bX_j-\bmu, \bV)$,
 where  {$d^2(\bx,\bv, \bSi)=(\bx-\bv)\trasp\bSi^{-1}(\bx-\bv)$} is the squared Mahalanobis distance.

Since $\bU \sim \itU(\itS_p)$, we have that $Y= (p-1)^{\frac 12} U_1 / \sqrt{1-U_1^2} \sim {\cal T}_{1,p-1}$
(see Muirhead,  1982, pp.38) and $U_1=Y/\sqrt{p-1+Y^2}$.
Moreover,   since $\bU\trasp\bv\sim U_1$ for any $\bv\in \itS_p$ we have that $\bU\trasp\bV_n^{-1/2} \left(\bX_i-\bX_j\right)\sim U_1 D_{ij}^{-}$ and $\bU\trasp\bV_n^{-1/2} \left(\bX_i+\bX_j-2\bm_n\right)\sim U_1 D_{ij}^{+}$. These facts entail that
 \begin{eqnarray*}
T_{n, \bm,\bV }
&=&\frac{1}{2\,n}  \sum_{i,j}  \left[f\left( b\, D_{ij}^{-} \right) - f\left( b \, D_{ij}^{+} \right)\right] \;,
\end{eqnarray*}
where the function $f:\real\to\real$ is defined as
$$
f(u)=\esp_{{\cal T}_{1,p-1}} \left( \frac{\sin\left(u\frac{Y}{\sqrt{p-1+Y^2}}\right)}{ \frac{Y}{\sqrt{p-1+Y^2}}u} \right)=\esp_{\itU_p} \left( \frac{\sin\left(u\,U_1\right)}{ u\,U_1} \right)\;,
$$
with $\sin(u)/u=1$ if $u=0$.
Then, using that $D_{ij}^{-}=D_{ji}^{-}$, $D_{ij}^{+}=D_{ji}^{+}$, $D_{ii}^{-}=0$ and $f(0)=1$, we get a simpler expression for the test statistic given by
$$
T_{n, \bm,\bV }=\frac{1}{2n} \sum_{j=1}^n\left\{1-f \left(bD_{jj}^{+}\right)+2 \sum_{i=1}^{j-1}\left[ f \left(bD_{ij}^{-}\right)-f \left(bD_{ij}^{+}\right)\right]\right\}\;.$$
To summarize, in order to compute the test statistic, the user only has to evaluate the function $f$ by Monte Carlo over a grid of points. 

{\setcounter{equation}{0}
\renewcommand{\theequation}{B.\arabic{equation}}
\section*{Appendix B: Proofs of Theorems \ref{main}.1 and  \ref{main}.2}{ \label{proof}}

\noi \textsc{Proof of Theorem \ref{main}.1.}
Using that $\sin(x-y)=\sin(x)\cos(y)-\cos(x)\sin(y)$ and since $\ba\trasp\bV_n^{-1/2}(\bX-\bm_n)= \ba\trasp\bV_n^{-1/2}(\bX-\bmu)-  \ba\trasp\bV_n^{-1/2}(\bm_n-\bmu)$, we have that $\sqrt{n}\,P_n\sin\left[t\ba\trasp\bV_n^{-1/2}(\bX-\bm_n)\right]=\sqrt{n}\,W_{1,n}(t,\ba)-\sqrt{n}\,W_{2,n}(t,\ba)$ where
\begin{eqnarray*}
W_{1,n}(t,\ba)&=&\cos\left[t\ba\trasp\bV_n^{-1/2}(\bm_n-\bmu)\right]\; P_n\sin\left[t\ba\trasp\bV_n^{-1/2}(\bX-\bmu)\right]\\
W_{2,n}(t,\ba)&=&\sin\left[t\ba\trasp\bV_n^{-1/2}(\bm_n-\bmu)\right]\; P_n\cos\left[t\ba\trasp\bV_n^{-1/2}(\bX-\bmu)\right]\;.
\end{eqnarray*}
Denote as $Z_n(t,\ba,\bA)=P_n\cos\left[t\ba\trasp\bA(\bX-\bmu)\right]$ and $\zeta(t,\ba,\bA)=\esp_P\cos\left[t\ba\trasp\bA(\bX-\bmu)\right]$. Note that since $\bX\sim\itE_p(\bmu, \bSi, \psi)$, we have that  $\zeta(t,\ba,\bA)= \psi(t^2\ba\trasp \bA\bSi \bA\trasp \ba)$ which entails that $\zeta(t,\ba,\bSi^{-1/2} )=\psi(t^2)$. The Dominated Convergence Theorem implies that 
$$   \lim_{ \bA \to \mbox{\small$\bSi^{-1/2}$}} \esp_P \sup_{t\in \real, \ba\in \itS_p}\left|\cos\left[t\ba\trasp\bA(\bX-\bmu)\right]-\cos\left(t\ba\trasp \bSi^{-1/2}(\bX-\bmu)\right)\right| =0\,, $$
which, together with the fact that $\bV_n\convprob \bSi$, entails that $\sup_{t\in \real, \ba\in \itS_p}|\zeta(t,\ba,\bV_n^{-1/2})-\zeta(t,\ba,\bSi^{-1/2})|\convprob 0$, that is, 
\begin{equation}
\sup_{t\in \real, \ba\in \itS_p}\left|\esp_P\cos\left[t\ba\trasp\bV_n^{-1/2}(\bX-\bmu)\right]-\psi(t^2)\right|\convprob 0\,.
\label{convesp}
\end{equation} 
Let $\|\cdot\|_{\ese}$ be a norm in the space $\itM$ of symmetric positive definite matrices. Then, as $\bV_n\convprob \bSi$, we have that for $n$ large enough with high probability, $\|\bV_n^{-1/2}-\bSi^{-1/2}\|_{\ese}\le 1$. Let $\itF=\{f(\bx)=\cos(t\ba\trasp\bA(\bx-\bmu)), \; t\in\itI, \ba \in \itS_p, \bA\in \real^{p\times p}: \|\bA-\bSi^{-1/2}\|_{\ese}\le 1\}$ and $\itG_C=\{f_\bb(\bx)= \cos(\bb\trasp(\bx-\bmu)), \; \bb\in \real^{p}: \|\bb\| \le C\}$. Then, for some $C>0$, $\itF\subset  \itG_C $. Note that the functions $f_\bb(\bx)$ are such that the map $\bb\mapsto f_\bb(\bx)$ is continuous for all fixed $\bx$ and its envelope $F(\bx)=\sup_{\|\bb\| \le C} |f_\bb(\bx)|$ satisfies that $F\in L_1(P)$ since $F\le 1$. Hence, using Lemma 3.10 in van de Geer (2000), we have that the class $\itG_C$ has finite bracketing number $N_{[\,]}(\epsilon, \itG_C, L_1(P))$, which entails that $\itG_C$ is a Glivenko--Cantelli class of functions, that is,  $\sup_{\|\bb\| \le C}|P_n\cos\left[\bb\trasp(\bX-\bmu)\right]-P \cos\left[\bb\trasp(\bX-\bmu)\right]|\convpp 0$. This convergence implies that
$\sup_{t\in\itI, \ba \in \itS_p}|Z_n(t,\ba,\bV_n^{-1/2})-\zeta(t,\ba,\bV_n^{-1/2})|\convprob 0$.
Hence, using (\ref{convesp}), we get that
\begin{equation}
\sup_{t\in\itI, \ba \in \itS_p}|Z_n(t,\ba,\bV_n^{-1/2})-\psi(t^2)|\convprob 0\,.
\label{convespcoseno}
\end{equation}
From the fact that $\lim_{u\to 0}\sin u/u=1$ , the consistency of $\bV_n$ and $\bm_n$, we get that
\begin{equation}
\sup_{t\in\itI, \ba \in \itS_p}\left|\frac{\sin\left[t\ba\trasp\bV_n^{-1/2}(\bm_n-\bmu)\right]}{t\ba\trasp\bV_n^{-1/2}(\bm_n-\bmu)}-1\right|\convprob 0\;.
\label{pasoseno}
\end{equation}
Using that $\sqrt{n}(\bm_n-\bmu)=O_\prob(1)$, together with (\ref{convespcoseno}) and (\ref{pasoseno}), we conclude that
$$\sup_{t\in\itI, \ba \in \itS_p} \left|\sqrt{n}\,W_{2,n}(t,\ba)- \psi(t^2)\left[t\ba\trasp\bSi^{-1/2}\sqrt{n}(\bm_n-\bmu)\right]\right|\convprob 0\;,$$
 which together with the fact that $\bm_n$ admits a Bahadur expansion leads to
\begin{equation}
\sqrt{n}\,W_{2,n}(t,\ba)=\psi(t^2)\;t\ba\trasp\bSi^{-1/2}\frac 1{\sqrt{n}} \sum_{i=1}^n \left(\bX_i-\bmu \right) \;\alpha_{\bm} \left(\| \bSi^{-1/2}\left(\bX_i-\bmu \right)\|\right)+R_{2,n}(t,\ba)\;,
\label{desarrolloW2}
\end{equation}
where $\sup_{t\in\itI, \ba \in \itS_p} |R_{2,n}(t,\ba)|\convprob 0$. 

Similarly, using that $|\cos(u)-1|\le |u|$, that $\itI$ is a bounded interval and the consistency of $\bV_n$ and $\bm_n$, we obtain that
\begin{equation}
\sup_{t\in\itI, \ba \in \itS_p}\left| \cos\left[t\ba\trasp\bV_n^{-1/2}(\bm_n-\bmu)\right]-1\right|\convprob 0\;.
\label{pasocoseno}
\end{equation}
Hence, the proof will be complete if we show that 
\begin{itemize}
\item[a)] $\sqrt{n}\,W_{3,n}(t,\ba)=\sqrt{n}P_n\sin\left[t\ba\trasp\bSi^{-1/2}(\bX-\bmu)\right]$ converges to a Gaussian process and
\item[b)] 
$
\sup_{t\in\itI, \ba \in \itS_p}\left|\sqrt{n}P_n\sin\left[t\ba\trasp\bV_n^{-1/2}(\bX-\bmu)\right]-\sqrt{n}P_n\sin\left[t\ba\trasp\bSi^{-1/2}(\bX-\bmu)\right]\right|\convprob 0\,.$
\end{itemize}
Effectively, if a) and b) hold the process $\sqrt{n}P_n\sin\left[t\ba\trasp\bV_n^{-1/2}(\bX-\bmu)\right]$ is tight, so, using (\ref{pasocoseno}), we can write
$
\sqrt{n}\,W_{1,n}(t,\ba)= \sqrt{n}\,W_{3,n}(t,\ba)+R_{1,n}(t,\ba)\;,
$
where $\sup_{t\in\itI, \ba \in \itS_p} |R_{1,n}(t,\ba)|\convprob 0$, which together with (\ref{desarrolloW2}) leads to
\begin{eqnarray*}
\sqrt{n}\,P_n\sin\left[t\ba\trasp\bV_n^{-1/2}(\bX-\bm_n)\right]&=&\psi(t^2)\;t\ba\trasp\bSi^{-1/2}\frac 1{\sqrt{n}} \sum_{i=1}^n \left(\bX_i-\bmu \right) \;\alpha_{\bm} \left(\| \bSi^{-1/2}\left(\bX_i-\bmu \right)\|\right)\\
&&+\sqrt{n}P_n\sin\left[t\ba\trasp\bSi^{-1/2}(\bX-\bmu)\right]+R_{n}(t,\ba)\;,
\end{eqnarray*}
where  $\sup_{t\in\itI, \ba \in \itS_p} |R_{n}(t,\ba)|\convprob 0$.

The proof of a) follows from  Ghosh and Ruymgaart (1992), so   it only remains to show b). 

To derive b), denote $Z_n^{\star}(t,\ba,\bA)= P_n\sin\left[t\ba\trasp\bA(\bX-\bmu)\right]$. Then, we have that $\esp_P Z_n^{\star}(t,\ba,\bA)=0$, since $\bX\sim \itE_p(\bmu, \bSi, \psi)$. Note that $\itF^{\star}=\{f(\bx)=\sin(t\ba\trasp\bA(\bx-\bmu)), \; t\in\itI, \ba \in \itS_p, \bA\in \real^{p\times p}: \|\bA-\bSi^{-1/2}\|_{\ese}\le 1\}\subset \itG_C^{\star}$ for some $C>0$ where $\itG_C^{\star}=\{f_\bb(\bx)= \sin(\bb\trasp\bSi^{-1/2}(\bx-\bmu)), \; \bb\in \real^{p}: \|\bb\| \le C\}$. In the Appendix C, it is shown   that  $\itG_C^{\star}$ is Donsker, which entails the uniform equicontinuity leading to b). 
 \square

\vskip0.2in
\noi \textsc{Proof Theorem \ref{main}.2.}  As in   Theorem \ref{main}.1, let $\itW_n(t,\ba)=\sqrt{n}P_n   \sin\left[ t\ba\trasp \bV_{n,\bX}^{-1/2} \left( \bX-\bm_{n,\bX} \right)   \right] $. We will show that the process $\itW_n=\left\{\itW_n(t,\ba)  \;, \;(t,\ba)\in \itI\times \itS_p\right\} $ converges in distribution to the Gaussian process $\itW^\star=\{\itW^\star(t,\ba)\;, \;(t,\ba)\in \itI\times \itS_p \} $ with $\itW^\star(t,\ba)=\itW(t,\ba)+({1}/{\ell!})B_\ell(t,\ba)$.

 As in the proof of Theorem \ref{main}.1, we have that  $\itW_n(t,\ba)=\sqrt{n}\,W_{1,n}(t,\ba)-\sqrt{n}\,W_{2,n}(t,\ba)$ where
\begin{eqnarray*}
W_{1,n}(t,\ba)&=&\cos\left[t\ba\trasp\bV_{n,\bX}^{-1/2}\left(\bm_{n,\bX}-\bmu\right)\right]\; P_n\sin\left[t\ba\trasp\bV_{n,\bX}^{-1/2}\left(\bX-\bmu)\right)\right]\\
W_{2,n}(t,\ba)&=&\sin\left[t\ba\trasp\bV_{n,\bX}^{-1/2}\left(\bm_{n,\bX}-\bmu)\right)\right]\; P_n\cos\left[t\ba\trasp\bV_{n,\bX}^{-1/2}\left(\bX-\bmu)\right)\right]\;.
\end{eqnarray*}
Besides, we also have that 
\begin{eqnarray}
\sup_{t\in\itI, \ba \in \itS_p}\left| \cos\left[t\ba\trasp\bV_{n,\bX}^{-1/2}\left(\bm_{n,\bX}-\bmu)\right)\right]-1\right|&\convprob& 0\;,
\label{pasocoseno2}\\
\sup_{t\in\itI, \ba \in \itS_p}\left|\frac{\sin\left[t\ba\trasp\bV_{n,\bX}^{-1/2}\left(\bm_{n,\bX}-\bmu)\right)\right]}{t\ba\trasp\bV_{n,\bX}^{-1/2}\left(\bm_{n,\bX}- \bmu)\right)}-1\right|&\convprob &0\;,
\label{pasoseno2}
\end{eqnarray}
 hold since $\bV_{n,\bX}\convprob \bSi$ and $\bm_{n,\bX}-\bmu\convprob 0$. On the other hand, from  (\ref{condicion_mux}) 	 and using that  $\bm_{n,\bZ}$ satisfies (\ref{vonmises}) together with the fact that $\bV_{n,\bX}\convprob \bSi$, we get that
 \begin{equation}
\sqrt{n}\bV_{n,\bX}^{-1/2}\left(\bm_{n,\bX}- \bmu\right)=\frac 1{\sqrt{n}} \sum_{i=1}^n  {\bSi^{-1/2}}\left(\bZ_i-\bmu \right) \;\alpha_{\bm} \left(\| \bSi^{-1/2}\left(\bZ_i-\bmu \right)\|\right)+R_n\,,
\label{seno2}
\end{equation}
where $R_n\convprob 0$.  Hence, if we show that
\begin{equation}
\sup_{t\in\itI, \ba \in \itS_p}\left|P_n\cos\left[t\ba\trasp\bV_{n,\bX}^{-1/2}\left(\bX-\bmu\right)\right]-\psi(t^2)\right|\convprob 0\,,
\label{coseno2}
\end{equation}
we obtain the following expansion for $\sqrt{n}\,W_{2,n}(t,\ba)$
\begin{eqnarray*}
\sqrt{n}\,W_{2,n}(t,\ba)&=&\psi(t^2)\;t\ba\trasp\bSi^{-1/2}\frac 1{\sqrt{n}} \sum_{i=1}^n \left(\bZ_i-\bmu \right) \;\alpha_{\bm} \left(\| \bSi^{-1/2}\left(\bZ_i-\bmu \right)\|\right)+R_{2,n}(t,\ba)\;,
 \end{eqnarray*}
where $\sup_{t\in\itI, \ba \in \itS_p} |R_{2,n}(t,\ba)|\convprob 0$.

To obtain (\ref{coseno2}), notice that from  the proof of Theorem \ref{main}.1, we have that
\begin{equation}
\sup_{t\in\itI, \ba \in \itS_p}|P_n\cos\left[t\ba\trasp\bV_{n,\bX}^{-1/2}(\bZ-\bmu)\right]-\psi(t^2)|\convprob 0\,,
\label{coseno3}
\end{equation}
since $\bZ\sim \itE_p(\bmu, \bSi, \psi)$ and $\bV_{n,\bX}\convprob \bSi$.
Besides, using that $|\cos(u)-\cos(v)|\le |u-v|$  and $X_i=\bZ_i+\bY_i/n^{\alpha}$,  we get the bound
\begin{eqnarray*}
&&\sup_{t\in\itI, \ba \in \itS_p}|P_n\cos\left[t\ba\trasp\bV_{n,\bX}^{-1/2}\left(\bX-\bmu\right)\right]-P_n\cos\left[t\ba\trasp\bV_{n,\bX}^{-1/2}(\bZ-\bmu)\right]| \le    \lambda_{n,\max}^{1/2}\, \frac{\nu}{n^\alpha}\, P_n \|\bY\|\,,
\end{eqnarray*}
where   $\lambda_{n,\max} $ stands for the largest eigenvalue of the   matrix $\bV_{n,\bX}^{-1}$. Therefore, using that $\esp \|\bY\|<\infty$ and $\bV_{n,\bX}\convprob \bSi$, we get (\ref{coseno2}) from (\ref{coseno3}).

Denote as    $M_{n,\bX}(t,\ba, \bA)=P_n\sin\left[t\ba\trasp\bA\left(\bX-\bmu\right)\right]$ and  $M_{n,\bZ}(t,\ba, \bA)=P_n\sin\left[t\ba\trasp\bA\left(\bZ-\bmu\right)\right]$. Then, noting that $W_{1,n}(t,\ba)=\cos\left[t\ba\trasp\bV_{n,\bX}^{-1/2}\left(\bm_{n,\bX}-\bmu\right)\right]
M_{n,\bX}(t,\ba, \bV_{n,\bX}^{-1/2})$, from (\ref{pasocoseno2}), we obtain that  
$\sqrt{n}\,W_{1,n}(t,\ba)=M_{n,\bX}(t,\ba, \bV_{n,\bX}^{-1/2})+R_{1,n}(t,\ba)$, 
with $\sup_{t\in\itI, \ba \in \itS_p} |R_{1,n}(t,\ba)|\convprob 0$, if 
\begin{equation}
\sqrt{n}\;\sup_{t\in\itI, \ba \in \itS_p} |M_{n,\bX}(t,\ba, \bV_{n,\bX}^{-1/2})|=O_\prob(1)
\label{tasaMn}
\end{equation}
holds.  

Recall that, from the proof of Theorem \ref{main}.1, $\sqrt{n} M_{n, \bZ}(t,\ba, \bV_{n,\bX}^{-1/2})$ converges    to a Gaussian process since $\bV_{n,\bX}\convprob \bSi$. 
Let $\delta$ be such that $\|\bA-\bSi^{-1/2}\|_{\ese}\le \delta$ entails that $|\lambda_{\max}(\bA\bSi\bA\trasp)-1|\le \delta_0/\nu$ where $\delta_0=\min_{1\le m\le \ell}(\delta_m)$ and $\delta_m$ are given in assumption f). Denote $\itF=\{f(\by,\bz)=(t\ba\trasp \bA \by  )^\ell\sin^{(\ell)}(t\ba\trasp\bA(\bz-\bmu)), \; ((t,\ba,\bA)\in \itA\}$, where $\itA =\{(t,\ba, \bA): t\in\itI, \ba \in \itS_p,   \|\bA-\bSi^{-1/2}\|_{\ese}\le \delta\}$. Then,  
 the proof will be completed if we show  the following convergences
 \begin{eqnarray}
 && \hskip-0.9in \sup_{(t,\ba,\bA)\in \itA} \left|\sqrt{n}\left\{M_{n,\bX}(t,\ba, \bA)-M_{n,\bZ}(t,\ba, \bA)\right\}-\frac{1}{\ell!} P_n \left\{ \left(t\ba\trasp \bA \bY \right)^\ell \sin^{(\ell)}(t\ba\trasp\bA(\bZ-\bmu))\right\}\right| \convprob 0\,,
\label{primera}\\
&& \sup_{f\in\itF} \left|  P_n f(\bY,\bZ)- P  f(\bY,\bZ)\right| \convprob 0\,,
\label{segunda}
\end{eqnarray} 
for some $\delta$ small enough, since  (\ref{primera}) and (\ref{segunda}) entail (\ref{tasaMn}) and also the desired expansion. 
For any $1\le s\le \ell$, define 
\begin{equation}
\itG_s=\{g_{\bb }(\by, \bz)= (\bb\trasp \bSi^{-1/2} \by  )^s\sin^{(s)}(\bb\trasp\bSi^{-1/2}(\bz-\bmu)), \; \bb\in \real^{p}:  \|\bb\| \le \nu+\delta_0\}\,.
\label{Gs}
\end{equation}
The proof of  (\ref{segunda}) follows using Lemma 3.10 of van de Geer (2000) and similar arguments to those considered in  proof of Theorem \ref{main}.1 applied to  the classes of functions  $\itF$ and $\itG_\ell$ since   $\itF\subset  \itG_\ell $ and the envelope $G(\by,\bz)=\sup_{ \|\bb\| \le C } |g_{\bb }(\by, \bz)|\le    C^\ell\|\by\|^\ell \in L^1(P)$ with $C=(\nu+\delta_0)\lambda_{\max}^{1/2}(\bSi^{-1})$.

It only remains to prove (\ref{primera}) which follow if we show that $ \sup_{\|\bb\|  \le  \nu+\delta_0} \left|Y_n(\bb)\right| \convprob 0$ where 
\begin{eqnarray*}
Y_n(\bb)&=&
 \sqrt{n}\left\{P_n\sin\left[\bb\trasp \bSi^{-1/2}\left(\bX-\bmu\right)\right]-P_n\sin\left[\bb\trasp \bSi^{-1/2}\left(\bZ-\bmu\right)\right] \right\}\\
&& -\frac{1}{\ell!} P_n \left\{ \left( \bb\trasp \bSi^{-1/2}  \bY \right)^\ell \sin^{(\ell)}(\bb\trasp \bSi^{-1/2}(\bZ-\bmu))\right\}\,.
\end{eqnarray*}
Using a Taylor's expansion and noting that $\bX_i-\bmu=(\bZ_i-\bmu)+ \bY_i/n^{\alpha}$ and $\alpha=1/(2\ell)$ we have that
$(1/n) \sum_{i=1}^n\sin\left[\bb\trasp \bSi^{-1/2}  \left(\bX_i-\bmu\right)\right]=(1/n) \sum_{i=1}^n\sin\left[\bb\trasp\bSi^{-1/2}  \left(\bZ_i-\bmu)\right)\right]+ S_{1,n}+S_{2,n}+S_{3,n}$,
where
\begin{eqnarray*}
S_{1,n}&=&\frac 1n \sum_{i=1}^n\sum_{s=1}^{\ell-1} \frac{1}{s!} \frac{1}{n^{\alpha\, s}}\left[\bb\trasp \bSi^{-1/2}  \bY_i \right]^s \sin^{(s)}(\bb\trasp \bSi^{-1/2} (\bZ_i-\bmu))=\sum_{s=1}^{\ell-1} \frac{1}{s!}\, \frac{1}{n^{\alpha\, s}}\, S_{1,n,s}\\
S_{2,n}&=&\frac{1}{n^{\alpha\, \ell}}\frac{1}{\ell!} \frac 1n \sum_{i=1}^n   \left( \bb\trasp   \bSi^{-1/2} \bY_i \right)^\ell \sin^{(\ell)}(\bb\trasp\bSi^{-1/2}  (\bZ_i-\bmu))\\
S_{3,n}&=& \frac{1}{n^{\alpha\, (\ell+1)}}\frac{1}{(\ell+1)!} \frac 1n \sum_{i=1}^n   \left( \bb\trasp  \bSi^{-1/2}  \bY_i \right)^{ \ell+1 }  \sin^{(\ell+1)}\left(\xi_n\right)
\end{eqnarray*}
with $\xi_n=\theta_n\bb\trasp \bSi^{-1/2} (\bZ_i-\bmu)+(1-\theta_n)\bb\trasp \bSi^{-1/2} \left(\bX_i-\bmu)\right)$, for some $\theta_n\in (0,1)$. Thus, $Y_n(\bb)= \sqrt{n}\left(S_{1,n}+S_{3,n}\right)$ so to conclude the proof, we only have to show that $\sqrt{n}\sup_{\|\bb\|  \le \delta_0+\nu}|S_{j,n}|\convprob 0$, for $j=1,3$.
 Using that  $\ell \alpha=1/2$,   we get that, for any $\|\bb\|\le \delta_0+\nu$,
\begin{eqnarray*}
|\sqrt{n}S_{3,n}|&\le & \frac{1}{n^{\alpha\,}}\frac{1}{(\ell+1)!} P_n \left|   \left( \bb\trasp  \bSi^{-1/2}  \bY \right)^{\ell+1 } \right| \le   \frac{1}{n^{\alpha\,}}\frac{1}{(\ell+1)!}   C^{\ell+1}\,P_n   \|\bY\|^{\ell+1} 
 \end{eqnarray*}
where  $C=(\nu+\delta_0)\lambda_{\max}^{1/2}(\bSi^{-1})$, which entails that  $\sqrt{n}S_{3,n}\convprob 0$ since $\esp\|\bY\|^{\ell+1}<\infty$.
 
To obtain that $\sqrt{n}\sup_{\|\bb\|  \le C}|S_{1,n}|\convprob 0$, it is enough to show that, for any $1\le s\le \ell-1$, $ \sqrt{n}\sup_{\|\bb\|  \le C}|S_{1,n,s}|/{n^{\alpha\, s}}\convprob 0$ which will follow if we prove that
\begin{equation}
 \sqrt{n}\sup_{\|\bb\|  \le C}|S_{1,n,s}|=O_\prob(1)\,.
\label{conv_S1ns}
\end{equation}
Note that by assumption f), for any $s<\ell$,  $\esp\left\{\left[\bb\trasp\bSi^{-1/2}  \bY_i \right]^s \sin^{(s)}(\bb\trasp\bSi^{-1/2} (\bZ_i-\bmu))\right\}=0$, when   $\|\bb\|\le  \delta_0+\nu$. Thus,  $\esp S_{1,n,s}=0$ holds, so    (\ref{conv_S1ns}) follows from the fact that the class $\itG_{s}$ defined in (\ref{Gs}) is   Donsker which is derived in the Appendix C. \square


{\setcounter{equation}{0}
\renewcommand{\theequation}{C.\arabic{equation}}

\section*{Appendix C: Proof that $\itG_C^{\star}=\{f_\bb(\bx)= \sin(\bb\trasp\bSi^{-1/2}(\bx-\bmu)), \; \bb\in \real^{p}: \|\bb\| \le C\}$ and $\itG_s$ defined in (\ref{Gs}) are Donsker}{\label{apenC}}
When $\esp_{P}\|\bX-\bmu\|^2 <\infty $,  the result follows easily from Lemma 2.5 in van de Geer (2000) and Theorem 2.7.11 in van der Vaart and Wellner (1996). However, since we do not assume this moment condition,  we have to work more carefully and we will use the fact that  $\int_0^1\sqrt{\log H_u}du <\infty$.

To provide a unified proof, denote as
$$\itG_s=\{g_{\bb }(\by, \bz)= (\bb\trasp \bSi^{-1/2} \by  )^s\sin^{(s)}(\bb\trasp\bSi^{-1/2}(\bz-\bmu)), \; \bb\in \real^{p}:  \|\bb\| \le C\}\,.$$
Hence, when $s=0$, $\itG_s=\itG_C^{\star}$ while for $C= \nu+\delta_0$ we get the class defined in (\ref{Gs}). It is then, enough to show that $\itG_s$ is Donsker when $\int_0^1\sqrt{\log H_u}du <\infty$ if $s=0$ or if $\bY$ and $\bZ$ are independent while, if $1\le s<\ell$ and $\bY$ and $\bZ$ are not independent we will use that for any $1\le s<\ell$,   $\int_0^{1} \sqrt{\log H_{u^{q}}}du<\infty$, where $q= {\ell}/({\ell-s})$,  $\sin^{(s)}(u)=(-1)^s \sin(u)$ or  $\sin^{(s)}(u)=(-1)^{s+1} \cos(u)$ and  that $\esp\|\bY\|^{2s}<\infty$.  

 For simplicity, denote  $\itB_p(\bb,\delta)=\{\bu: \|\bu-\bb\|\le \delta\}$, $\itB_p(\delta)= \itB_p(\bcero,\delta)$, $\|f\|_{L^2(P)}=(\esp_P f^2)^{1/2}$,   $\lambda=\sqrt{\lambda_{\max}(\bSi^{-1})}$ and $A=2\lambda^{s}\left\{3\left(\esp\|\bY\|^{2s}\right)^{1/2}  +C^s\left(\esp\|\bY\|^{2\ell}\right)^{\frac s{2\ell}}\right\}$.

For any fixed $\rho>0$  and $\bbe\in \itB_p(C)$, define $\Psi(\by,\bz,\bbe,\rho)=\sup_{ \{ \bb  \in \itB_p(C) \cap \itB_p(\bbech,\rho)\}} |g_{\bb}(\by, \bz)-g_{\bbech}(\by, \bz)| $.
Note that the continuity of the sinus entails that the supremum can be taken over $\qu^p$, so that  $\Psi(\bx,\bbe,\rho)$ is measurable for each $\bbe$ and $\rho>0$. Note that $|\sin^{(s)}(u)-\sin^{(s)}(v)|\le |u-v|$ entails that 
$
|g_{\bb_1}(\by, \bz)-g_{\bb_2}(\by, \bz)|\le \lambda^s \|\by\|^s \max\left(\|\bb_1-\bb_2\|, \|\bb_1-\bb_2\|^s\right)\left\{1+C^s\|\bSi^{-1/2}(\bz-\bmu)\|\right\}\,.$
 Then, if $S=\|\bSi^{-1/2}(\bZ-\bmu)\|$ we obtain the bound
\begin{equation}
\esp_P\Psi^2(\bY,\bZ,\bbe,\rho)\indica_{[0,M]}(S) \le \esp\|\bY\|^{2s}\lambda^{2s} \left\{1+C^sM\right\}^2  \max\left(\rho^2, \rho^{2s} \right)\;.
\label{cota2}
\end{equation}
On the other hand, since $
|g_{\bb_1}(\by, \bz)-g_{\bb_2}(\by, \bz)|\le \lambda^s \|\by\|^s  \left\{\|\bb_1-\bb_2\|^s+2C^s\right\}$, we have that \linebreak$\Psi^2(\bY,\bZ,\bbe,\rho)\le \lambda^{2s}  \left\{\rho^{2s}+2C^s\right\}^2\|\bY\|^{2s}$  and so, using the dominated convergence Theorem we get that $\esp_P\Psi^2(\bY,\bZ,\bbe,\rho)\to 0$ as $\rho\to 0$.

For a given $0<\epsilon<\min(1,A)$, let $\eta= \epsilon/A$, choose  $M_\epsilon$ as the smallest value such that $\prob(S>M_\epsilon)=\prob(\|\bSi^{-1/2}(\bX-\bmu)\|> M_{\epsilon})\le (\epsilon/A)^{2\,\ell/(\ell-s)} $ and define $\rho_\epsilon=\min\{\eta,\eta/[C^s M_\epsilon]\}<1$. Then, since $\max\left(\rho_\epsilon^2, \rho_\epsilon^{2s} \right)=\rho_\epsilon^2$, we get that
\begin{eqnarray}
\esp_P\Psi^2(\bY,\bZ,\bbe,\rho_\epsilon)\indica_{[0,M_\epsilon]}(S) &\le& \esp\|\bY\|^{2s}\lambda^{2s} \left\{\min\left(\eta,\frac{\eta}{C^s M_\epsilon}\right)+C^sM_\epsilon\;\min\left(\eta,\frac{\eta}{C^s M_\epsilon}\right)\right\}^2    \nonumber\\
&\le & \esp\|\bY\|^{2s}\lambda^{2s} 4 \eta^2
\label{cota3}
\end{eqnarray}
Let $N=N(\rho_\epsilon,\itB_p(C))$ the minimum number of balls of radius $\rho_\epsilon$ and center  in $\itB_p(C)$, needed to cover  the set $\itB_p(C)$. Then, $N$ is at most twice the number of balls  of radius $\rho_\epsilon$  needed to cover  the set  $\itB_p(C)$ for which a bound is given in Lemma 2.5 in van de Geer (2000). Hence, there exist $\bb_1,\dots, \bb_N$,  $\bb_j\in \itB_p(C) $    such that $\itB_p(C)\subset \cup_{j=1}^N\itB_p(\bb_j,\rho_\epsilon)$.
Define 
$
u_j(\by, \bz)=g_{\bb_j}(\by, \bz)+\Psi(\by, \bz,\bb_j,\rho_\epsilon) $ and $
\ell_j(\by, \bz)=g_{\bb_j}(\by, \bz)-\Psi(\by, \bz,\bb_j,\rho_\epsilon) $.
Then,    for any $\bb\in \itB_p(C)$, there exists $1\le j\le N$ such that $\bb\in \itB_p(\bb_j,\rho_\epsilon)$, so that 
$|g_{\bb}(\by, \bz)-g_{\bb_j}(\by, \bz)|\le \Psi(\by, \bz,\bb_j,\rho_\epsilon)$ which entails that $\ell_j(\by, \bz)\le g_{\bb}(\by, \bz)\le u_j(\by, \bz)$. On the other hand, $\Psi(\by, \bz,\bb_j,\rho_\epsilon)=\Psi_{1,j}(\by, \bz)+\Psi_{2,j}(\by, \bz)$ where 
$\Psi_{1,j}(\by, \bz)= \Psi(\by, \bz,\bb_j,\rho_\epsilon)\indica_{[0,M_\epsilon]}(\|\bSi^{-1/2}(\bx-\bmu)\|)$ and $ \Psi_{2,j}(\by, \bz)= \Psi(\by, \bz,\bb_j,\rho_\epsilon)\indica_{(M_\epsilon,+\infty)}(\|\bSi^{-1/2}(\bx-\bmu)\|)$.
 Note that  using (\ref{cota3}), we get that 
 $$\|\Psi_{1,j}\|_{L^2(P)} \le 2\left(\esp\|\bY\|^{2s}\right)^{1/2}\lambda^{s} \,\eta= 2\left(\esp\|\bY\|^{2s}\right)^{1/2}\lambda^{s} \,\frac{\epsilon}{A}\,.$$ 
 On the other hand, the fact that $\Psi^2(\by, \bz,\bb_j,\rho_\epsilon)\le \lambda^{2s} \left\{\rho_\epsilon^{s} +2C^s\right\}^2\|\bY\|^{2s} $ entails that  
 \begin{eqnarray}
 \|\Psi_{2,j}\|_{L^2(P)}&\le&  \lambda^{s}  \left\{\rho_\epsilon^{s} +2C^s\right\} \left(\esp \|\bY\|^{2s}  \indica_{S> M_\epsilon} \right)^{1/2}\nonumber\\
 &\le& \left(\esp\|\bY\|^{2s}\right)^{1/2}\lambda^{s} \,\eta +2\lambda^{s}C^s\left\{ \esp \left(\|\bY\|^{2s}  \indica_{S> M_\epsilon}\right) \right\}^{1/2}\label{queuso}
 \end{eqnarray}
 We will apply H\"older inequality with $p=\ell/s$, so that $1/p=s/\ell$ and $q=\ell/(\ell-s)$. Then
 $$ \esp \|\bY\|^{2s}  \indica_{S> M_\epsilon}\le  \left(\esp\|\bY\|^{2sp}\right)^{\frac 1p}  \left(\esp\indica_{S> M_\epsilon}\right)^{\frac 1q}=  \left\{\esp\|\bY\|^{2\ell}\right\}^{\frac s\ell} \left\{\prob\left(S> M_\epsilon\right) \right\}^{\frac{(\ell-s)}\ell}$$
 so, using that $\eta=\epsilon/A$ we get
  \begin{eqnarray*}
 \|\Psi_{2,j}\|_{L^2(P)} 
 &\le& \left(\esp\|\bY\|^{2s}\right)^{1/2}\lambda^{s} \,\frac{\epsilon}{A} +2\lambda^{s}C^s  \left\{\esp\|\bY\|^{2\ell}\right\}^{\frac s{2\ell}}
\left\{\prob\left(S> M_\epsilon\right) \right\}^{\frac{(\ell-s)}{2\ell}} \\
 &\le& \left\{\left(\esp\|\bY\|^{2s}\right)^{1/2}\lambda^{s} +2\lambda^{s}C^s\left\{\esp\|\bY\|^{2\ell}\right\}^{\frac s{2\ell}}\right\}\,\frac{\epsilon}{A}
 \end{eqnarray*}
 so that
\begin{eqnarray*}
\|u_j -\ell_j \|_{L^2(P)}&=&2\|\Psi_{1,j}+\Psi_{2,j}\|_{L^2(P)}\le 2\|\Psi_{1,j}\|_{L^2(P)}+2\|\Psi_{2,j}\|_{L^2(P)}\\
&\le & \left\{6\left(\esp\|\bY\|^{2s}\right)^{1/2}\lambda^{s} +2\lambda^{s}C^s\left\{\esp\|\bY\|^{2\ell}\right\}^{\frac s{2\ell}}\right\}\frac{\epsilon}{A}=\epsilon\,.
\end{eqnarray*}
Summarizing, we have shown that the bracketing number  $N_{[\,]}(  \epsilon, \itG_s, L^2(P))$ is smaller or equal than $N(\rho_\epsilon,\itB_p(C))$ which, from Lemma 2.5 in van de Geer (2000), may be bounded as
 $$N(\rho_\epsilon,\itB_p(C))\le 2\,\left(\frac{4C+\rho_\epsilon}{\rho_\epsilon}\right)^p\;.$$ 
  Note that if $\rho_\epsilon>C$, $N(\rho_\epsilon,\itB_p(C))\le 2\times 5^p$, otherwise 
$N(\rho_\epsilon,\itB_p(C))\le   2\, \left({5C}/{\rho_\epsilon}\right)^p$.
Thus,
\begin{equation}
\label{cotabracketing}
N_{[\,]}( \epsilon, \itG_s, L^2(P))\le  2\,\max\left(5^p,  \left[ \frac{5C}{\rho_\epsilon}\right]^p\right)\le 2\,\max\left(5^p, \left[ \frac{5C}{\epsilon}\right]^p, \left[ 5\, C^{s+1} A\, M_\epsilon\,\frac{1}{\epsilon}\right]^p\right)\,.
\end{equation}
Note that in (\ref{cotabracketing}) we can always assume that $M_\epsilon\ge 1$, otherwise we take $M_\epsilon=1$ which gives an upper bound, so that  $M_\epsilon=H_{(\epsilon/A)^{\ell/(\ell-s)} }$. Thus, if we denote as $A=p\log(5)+p\log(5\,C)+p\log(5\, C^{s+1} A)+\log(2)$, we have the following bound 
\begin{eqnarray*}
\int_0^1\sqrt{\log\left(N_{[\,]}(u, \itG_s, L^2(P))\right)} du&\le & \int_0^1 \sqrt{A+p\log M_u+2p\log\left(\frac{1}u\right)} du\\
&\le &  \sqrt{A}+\sqrt{p}\int_0^1 \sqrt{\log M_u}\; du +\sqrt{2p}\int_0^1\sqrt{\log\left(\frac{1}u\right)} du<\infty \,,
\end{eqnarray*}
since by hypothesis  $\int_0^1 \sqrt{\log M_u}du= \int_0^{1} \sqrt{\log H_{ (u/A)^{q}}}du=A\int_0^{A} \sqrt{\log H_{u^{q}}}du<\infty$,
where $q= {\ell}/({\ell-s})$, concluding the proof.

\vskip0.1in

Note that  if $s=0$, the condition $\int_0^{1} \sqrt{\log H_{u}}du<\infty$ suffices to prove that $ \itG_s$ is Donsker. Furthermore, if $\bY_i$ y $\bZ_i$  are independent the assumption $\int_0^{1} \sqrt{\log H_{u^{q}}}du<\infty$ is also weakened to $\int_0^{1} \sqrt{\log H_{u}}du<\infty$. Indeed, in this case, we define $A=2\lambda^{s}\left(\esp\|\bY\|^{2s}\right)^{1/2}\left\{3  +C^s\right\}$ and we choose  $M_\epsilon$ as the smallest value greater or equal than 1 such that $\prob(S>M_\epsilon)=\prob(\|\bSi^{-1/2}(\bX-\bmu)\|> M_{\epsilon})\le \epsilon^2/A^2 $, that is, $M_\epsilon=H_{(\epsilon/A)}$ and  using the independence in (\ref{queuso}), we get
 \begin{eqnarray*}
 \|\Psi_{2,j}\|_{L^2(P)}&\le&   \left(\esp\|\bY\|^{2s}\right)^{1/2}\lambda^{s} \,\frac{\epsilon}{A} +2\lambda^{s}C^s \left(\esp \|\bY\|^{2s} \;\prob\left(S> M_\epsilon\right) \right)^{1/2}
 \\
 &\le& \left\{\left(\esp\|\bY\|^{2s}\right)^{1/2}\lambda^{s} +2\lambda^{s}C^s\left(\esp\|\bY\|^{2s}\right)^{1/2}\right\}\,\frac{\epsilon}{A}\,,
 \end{eqnarray*}
 so that $
\|u_j -\ell_j \|_{L^2(P)}\le \epsilon$. So, as above,   the bracketing number  $N_{[\,]}(  \epsilon, \itG_s, L^2(P))$ is smaller or equal than $N(\rho_\epsilon,\itB_p(C))$ getting the bound (\ref{cotabracketing}). Hence,  if we denote as $A=p\log(5)+p\log(5\,C)+p\log(5\, C^{s+1} A)+\log(2)$, we obtain
\begin{eqnarray*}
\int_0^1\sqrt{\log\left(N_{[\,]}(u, \itG_s, L^2(P))\right)} du
&\le &  \sqrt{A}+\sqrt{p}\int_0^1 \sqrt{\log M_u}\; du +\sqrt{2p}\int_0^1\sqrt{\log\left(\frac{1}u\right)} du<\infty 
\end{eqnarray*}
since by hypothesis  $\int_0^1 \sqrt{\log M_u}du= \int_0^{1} \sqrt{\log H_{(u/A)}}du=A\int_0^{A} \sqrt{\log H_{u}}du<\infty$.

\begin{sidewaysfigure}
\centering
\small \hskip0.7in $n=20$ \hskip1.2in $n=50$\hskip1.2in $n=100$\hskip1.2in $n=200$\\
\includegraphics[scale=0.90]{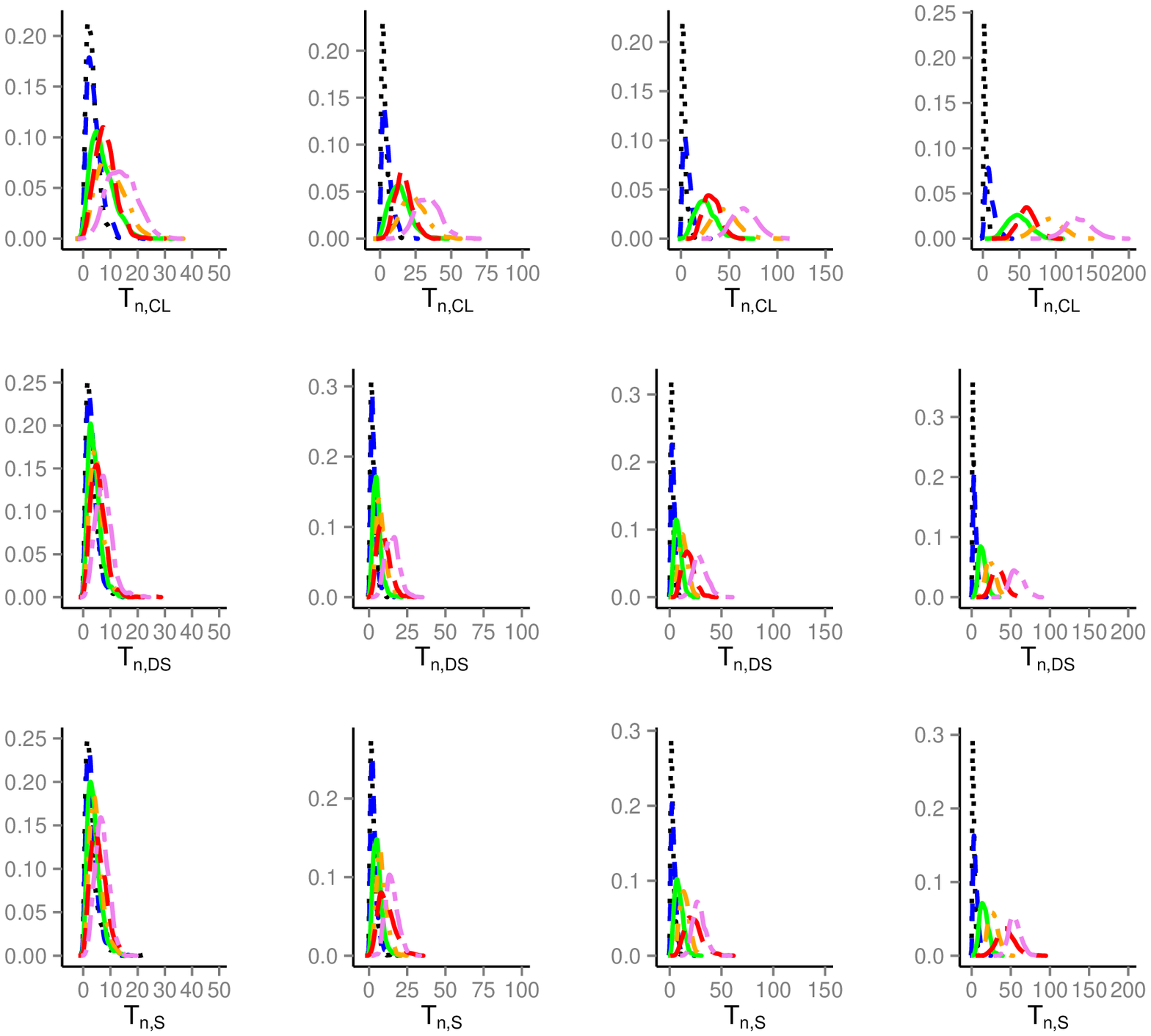}
\caption{\label{fig:densidadC0}{Density estimates of $T_{n,\ml}$, $T_{n,\ds}$ and $T_{n,\ese}$. Black and dotted line corresponds to $H_0^{(1)}$, blue  and dash line to $H_{1, 0.5}^{(1)}$, green and solid line to $H_{1,1}^{(1)}$, orange and dot--dashed line to $H_{1,1.5}^{(1)}$, red and long--dashed line to $H_1^{\star^{(1)}}$ and violet and two--dashed line to $H_1^{\star^{(2)}}$. }} 
\end{sidewaysfigure}

\begin{sidewaysfigure}
\centering
\small \hskip0.7in $n=20$ \hskip1.2in $n=50$\hskip1.2in $n=100$\hskip1.2in $n=200$\\
\includegraphics[scale=0.90]{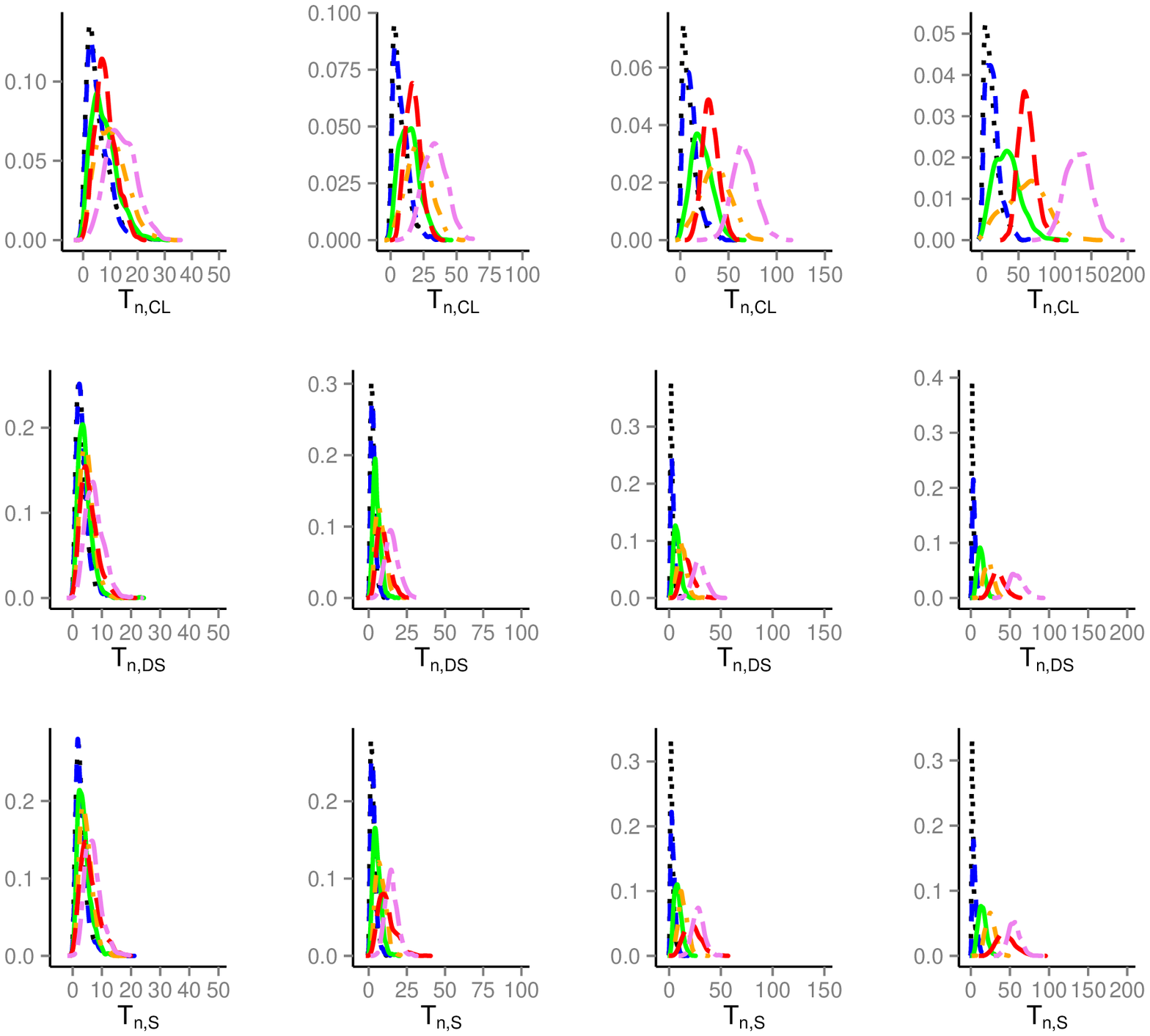}
\caption{\label{fig:densidadC1}{Density estimates of $T_{n,\ml}$, $T_{n,\ds}$ and $T_{n,\ese}$. Black and dotted line corresponds to $H_0^{(2)}$,  blue  and dash line to $H_{1, 0.5}^{(2)}$, green and solid line to $H_{1,1}^{(2)}$, orange and dot--dashed line to $H_{1,1.5}^{(2)}$, red and long--dashed line to $H_1^{\star^{(1)}}$ and violet and two--dashed line to $H_1^{\star^{(2)}}$. }} 
\end{sidewaysfigure}

\begin{sidewaysfigure}
\centering
\small \hskip0.7in $n=20$ \hskip1.2in $n=50$\hskip1.2in $n=100$\hskip1.2in $n=200$\\
\includegraphics[scale=0.90]{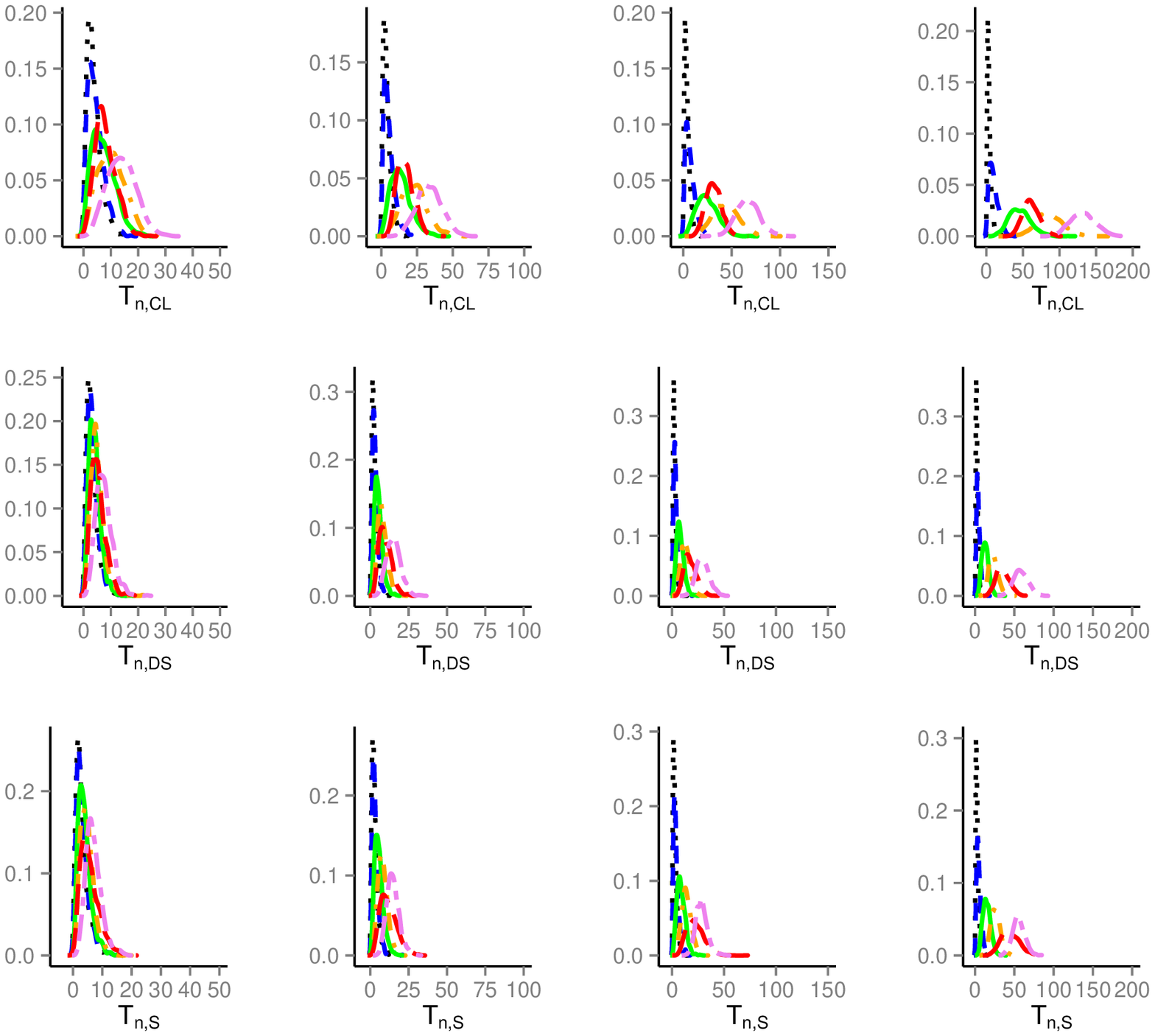}
\caption{\label{fig:densidadC3}{Density estimates of $T_{n,\ml}$, $T_{n,\ds}$ and $T_{n,\ese}$. Black and dotted line corresponds to $H_0^{(3)}$, blue  and dash line to $H_{1, 0.5}^{(3)}$, green and solid line to $H_{1,1}^{(3)}$, orange and dot--dashed line to $H_{1,1.5}^{(3)}$, red and long--dashed line to $H_1^{\star^{(1)}}$ and violet and two--dashed line to $H_1^{\star^{(2)}}$.}} 
\end{sidewaysfigure}

\begin{sidewaysfigure}
\centering
\small \hskip0.7in $n=20$ \hskip1.2in $n=50$\hskip1.2in $n=100$\hskip1.2in $n=200$\\
\includegraphics[scale=0.90]{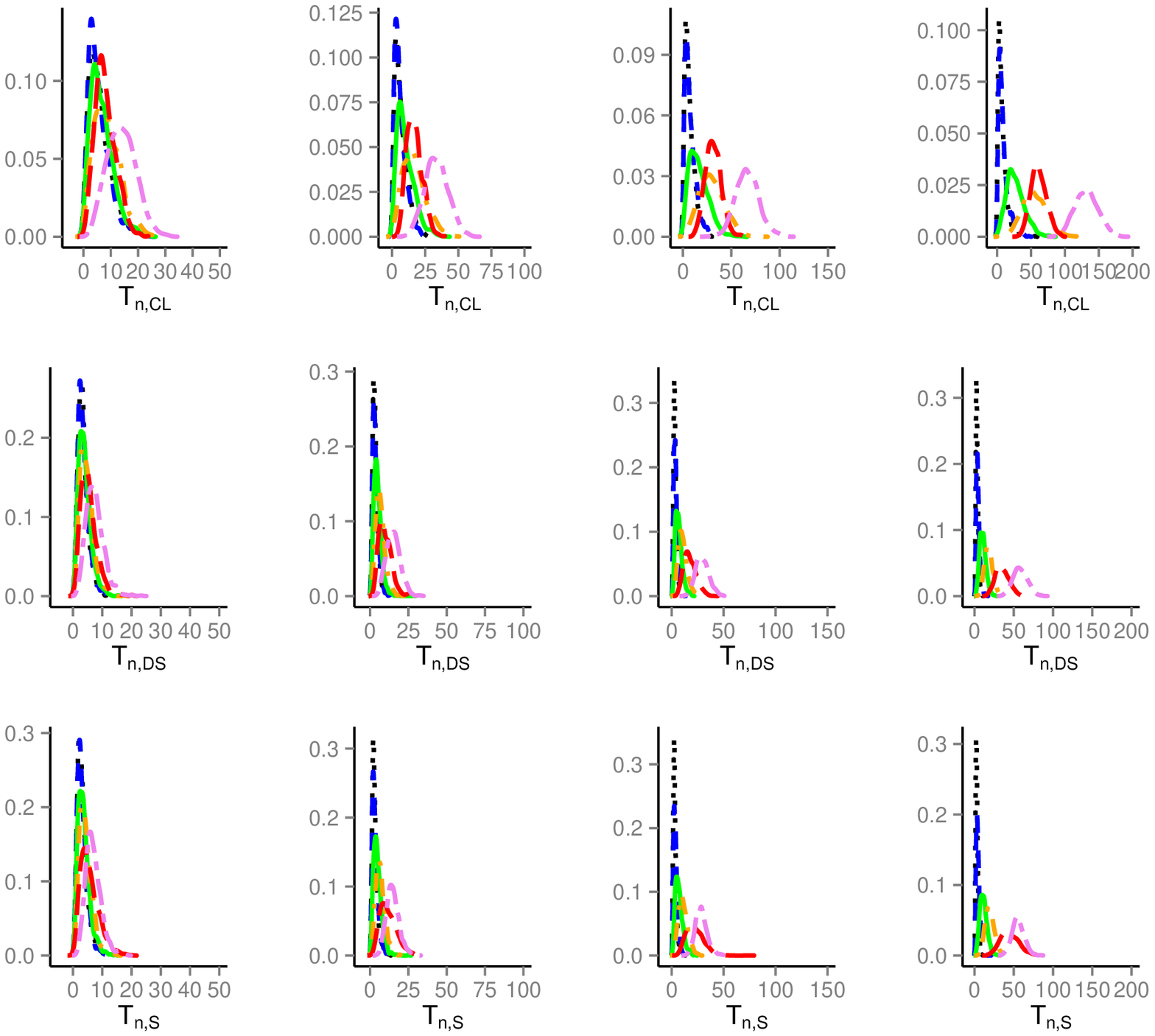}
\caption{\label{fig:densidadC4}{Density estimates of $T_{n,\ml}$, $T_{n,\ds}$ and $T_{n,\ese}$. Black and dotted line corresponds to $H_0^{(4)}$, blue  and dash line to $H_{1, 0.5}^{(4)}$, green and solid line to $H_{1,1}^{(4)}$, orange and dot--dashed line to $H_{1,1.5}^{(4)}$, red and long--dashed line to $H_1^{\star^{(1)}}$ and violet and two--dashed line to $H_1^{\star^{(2)}}$.}} 
\end{sidewaysfigure}

\begin{sidewaysfigure}
\centering
\small \hskip0.7in $n=20$ \hskip1.2in $n=50$\hskip1.2in $n=100$\hskip1.2in $n=200$\\
\includegraphics[scale=0.90]{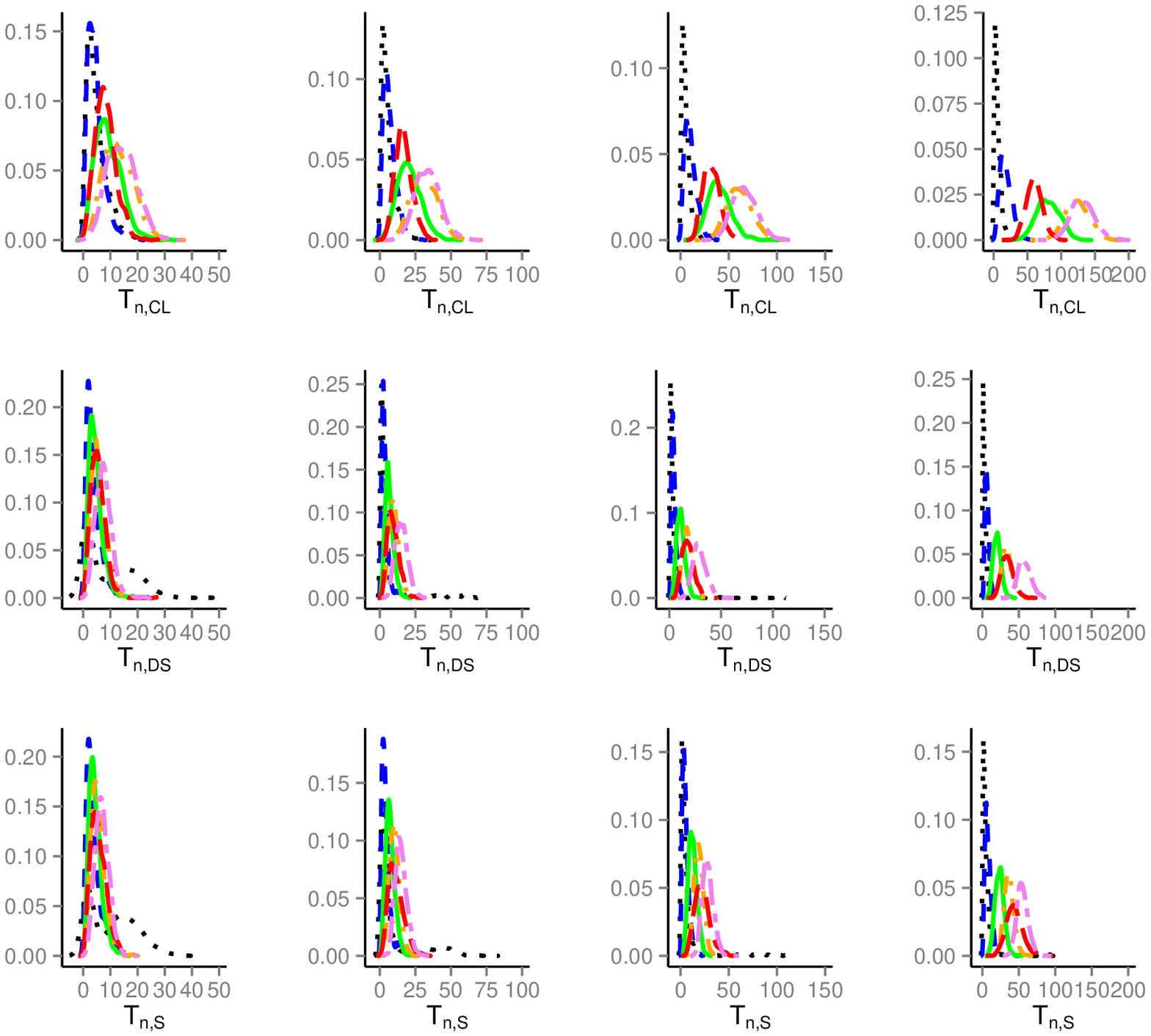}
\caption{\label{fig:densidadC5}{Density estimates of $T_{n,\ml}$, $T_{n,\ds}$ and $T_{n,\ese}$. Black and dotted line corresponds to $H_0^{(5)}$, blue  and dash line to $H_{1, 0.5}^{(5)}$, green and solid line to $H_{1,1}^{(5)}$, orange and dot--dashed line to $H_{1,1.5}^{(5)}$, red and long--dashed line to $H_1^{\star^{(1)}}$ and violet and two--dashed line to $H_1^{\star^{(2)}}$.}} 
\end{sidewaysfigure}

 \begin{sidewaysfigure}
\centering
\small \hskip0.7in $n=20$ \hskip1.2in $n=50$\hskip1.2in $n=100$\hskip1.2in $n=200$\\
\includegraphics[scale=0.90]{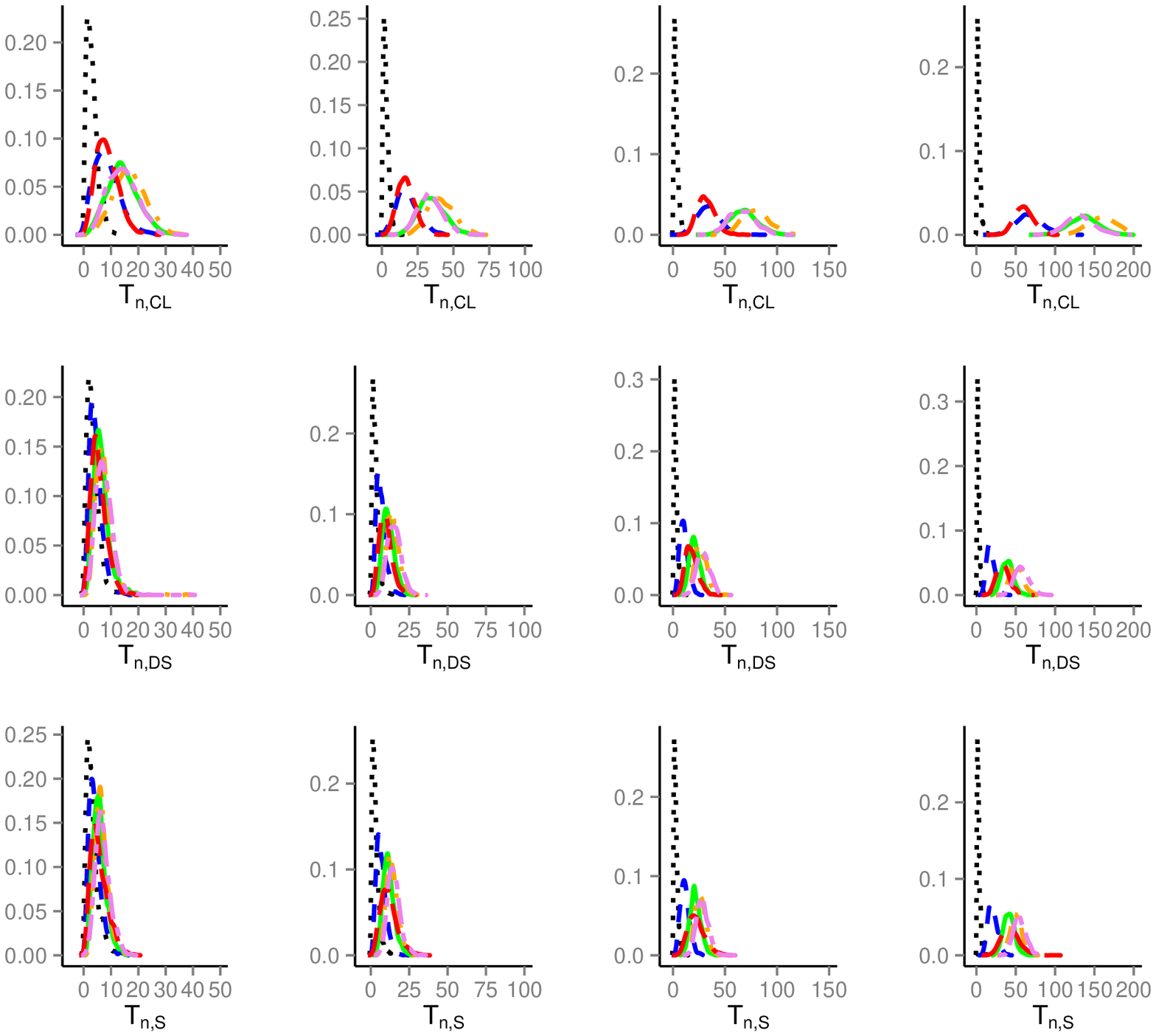}
\caption{\label{fig:densidadC6}{Density estimates of $T_{n,\ml}$, $T_{n,\ds}$ and $T_{n,\ese}$. Black and dotted line corresponds to $H_0^{(6)}$, blue  and dash line to $H_{1, 0.5}^{(6)}$, green and solid line to $H_{1,1}^{(6)}$, orange and dot--dashed line to $H_{1,1.5}^{(6)}$, red and long--dashed line to $H_1^{\star^{(1)}}$ and violet and two--dashed line to $H_1^{\star^{(2)}}$.}} 
\end{sidewaysfigure}

\begin{sidewaysfigure}
\centering
\small \hskip0.7in $n=20$ \hskip1.2in $n=50$\hskip1.2in $n=100$\hskip1.2in $n=200$\\
\includegraphics[scale=0.90]{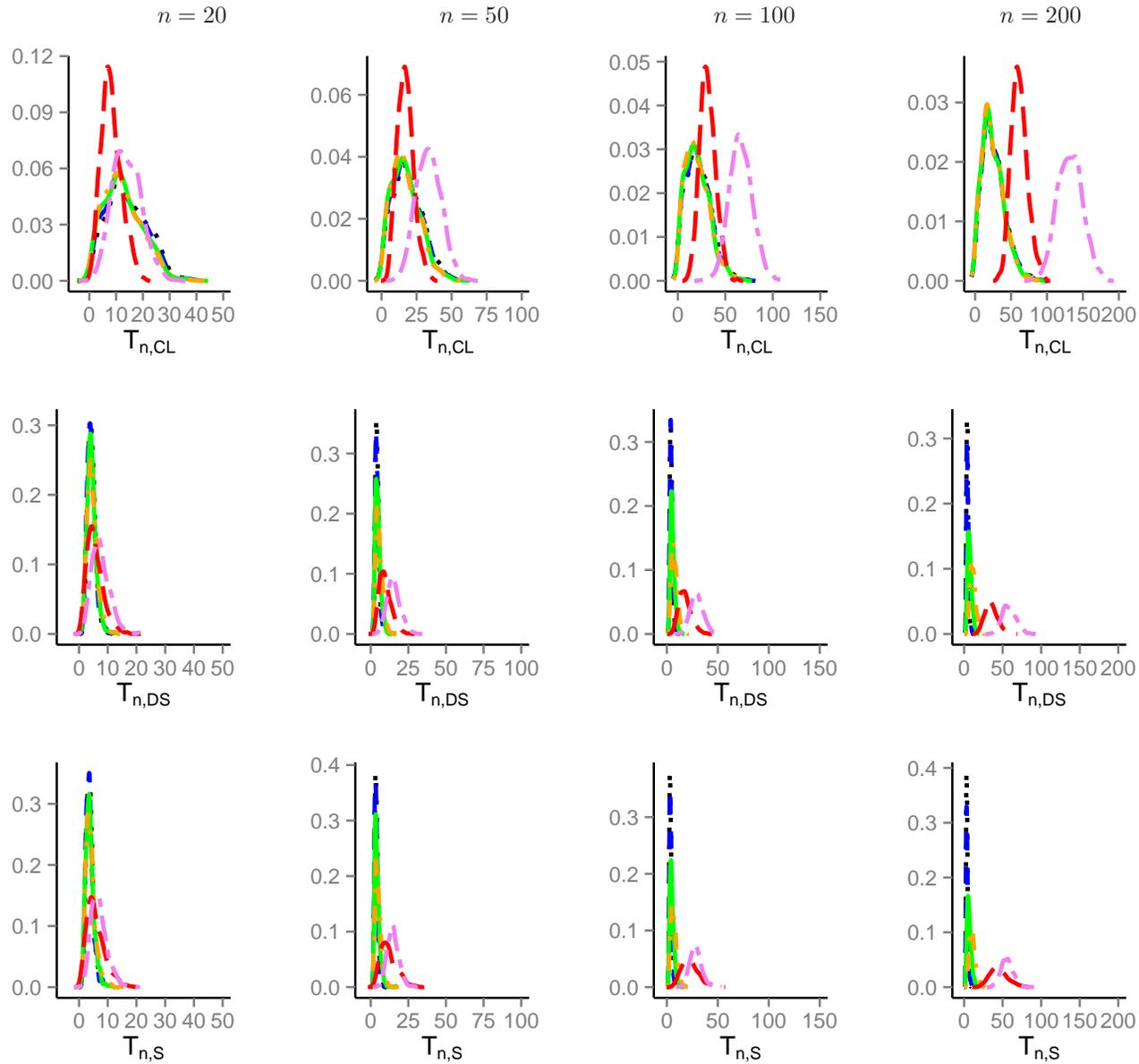}
\caption{\label{fig:densidadC2}{Density estimates of $T_{n,\ml}$, $T_{n,\ds}$ and $T_{n,\ese}$. Black and dotted line corresponds to $H_0^{(7)}$, blue  and dash line to $H_{1, 0.5}^{(7)}$, green and solid line to $H_{1,1}^{(7)}$, orange and dot--dashed line to $H_{1,1.5}^{(7)}$, red and long--dashed line to $H_1^{\star^{(1)}}$ and violet and two--dashed line to $H_1^{\star^{(2)}}$.}} 
\end{sidewaysfigure}

\small
\section*{References}
\footnotesize
\begin{description}

\item Anderson, T. W., Fang, K. T. and Hsu, H., 1986. Maximum--likelihood estimates and likelihood--ratio criteria for multivariate
 elliptically contoured distributions. \textsl{Canad. J. Statist.}, \textbf{14}, 55--59. 
  
\item Baringhaus, L., 1991. Testing for spherical symmetry of a multivariate distribution. \textsl{Ann. Statist.}, \textbf{19}, 899-917.

\item Batsidis, A. and Zografos, K., 2013. A necessary test of fit of specific elliptical distributions based on an estimator of Songs measure. \textsl{J. Multivariate Anal.}, \textbf{113}, 91--105.

\item Batsidis, A., Martin, N., Pardo, L. and Zografos, K., 2014. A necessary power divergence--type family of tests for testing elliptical symmetry. \textsl{J. Stat. Comput. Sim.}, \textbf{84}, 57--83.

\item Beran, R., 1979. Testing for elliptical symmetry of a multivariate density. \textsl{Ann. Statist.}, \textbf{7}, 150-162.

\item  Ghosh, S. and  Ruymgaart, F.H., 1992. Applications of empirical characteristic functions in some multivariate
problem. \textsl{Canad. J. Statist.},  \textbf{20},  429--440.

\item Fang, K. T. and Anderson, T. W. (eds.), 1990. \textsl{Statistical inference in elliptically contoured and related distributions}. Allerton Press, New York.

\item Fang, K. T., Kotz, S. and Ng, K. W., 1990. Symmetric multivariate and related distributions. \textsl{Monographs on Statistics and Applied Probability}, \textbf{36}, Chapman and Hall, London.

\item Fang, K.T., Zhu, L.X. and Bentler, P.M., 1993. A necessary test for sphericity of a high-dimensional distribution. \textsl{J. Multivariate Anal.}, \textbf{44}, 34-55.

\item Fernholz, L., 1983. \textsl{Von Mises calculus for statistical functionals}. Lecture Notes in Statistics, \textbf{19}, Springer Verlag, New York.

\item Hampel, F.R., Ronchetti, E.M.,  Rousseeuw, P.J. and  Stahel, W.A., 1986. \textsl{Robust Statistics: The Approach Based on Influence Functions},  Wiley, New York.

\item Huffer, F. and Park, C., 2007. A test for elliptical symmetry. \textsl{J. Multivariate Anal.}, \textbf{98}, 256--281.

\item Koltchinskii, V. and Li, L., 1998. Testing for spherical symmetry of a
multivariate distribution. \textsl{J. Multivariate Anal.}, \textbf{65}, 228--244.

\item Koltchinskii, V. and Sakhanenko, L., 2000. Testing for ellipsoidal symmetry of a
multivariate distribution. In: \textsl{High Dimensional Probability II}, Eds. Gin\'e, E., Mason, D. and Wellner, J., pp. 493--510.

\item Lopuha\"a, H., 1989. On the relation between $S-$estimators and $M-$estimators of multivariate location and covariance. \textsl{Annals of Statistics}, \textbf{17}, 1662--1683.

\item  Morales, D. ,  Pardo, L. , Pardo, M. C. and  Vajda, I., 2004. R\'enyi statistics for testing composite hypotheses in general exponential models. \textsl{Statistics}, \textbf{38}, 133--147.

\item Muirhead, R. J., 1982. \textsl{Aspects of Multivariate Statistical Theory}. John Wiley \& Sons, Canada.

\item Schott, J. R., 2002. Testing for elliptical symmetry in covariance-matrix-based analyses.\textsl{Statist. Probab. Lett.}, \textbf{60}, 395-404.

\item Tyler, D., 1982. Radial estimates and the test for sphericity. \textsl{Biometrika}, \textbf{69}, 429-436.

\item Ushakov, Nikolai G., 1999. \textsl{Selected Topics in Characteristic Functions}. Series: Modern Probability and Statistics, Walter de Gruyter.

\item van der Geer, S., 2000. \textsl{Empirical Processes in $M-$Estimation}. Cambridge University Press.

\item van der Vaart, A. and Wellner, J., 1996. \textsl{Weak Convergence and Empirical Processes. With Applications to Statistics}. New York: Springer.

\item Zhu, L.-X. and Neuhaus, G., 2003. Conditional tests for elliptical symmetry. \textsl{J. Multivariate Anal.}, \textbf{84},  284--298.

\end{description}

\end{document}